\documentclass[review,10pt]{elsarticle}

\usepackage{epsfig}
\usepackage{bm}
\usepackage{hyperref}
\usepackage{amsmath,amssymb,amsfonts,amsthm,bm}
\usepackage{cancel}
\usepackage{float}
\usepackage{miller}
\usepackage{subcaption}
\usepackage[margin=1in]{geometry}
\usepackage{algpseudocode}
\usepackage{soul}
\usepackage{outlines}
\usepackage{todonotes}
\usepackage{enumitem}
\biboptions{sort&compress}
\usepackage[authormarkup=none]{changes}

\usepackage{color}

\definecolor{ic}{HTML}{FF4047}
\definecolor{ea}{HTML}{FF6347}

\newcommand\U{\langle\bm{u}\rangle}

\newtheorem{definition}{Definition}

\definechangesauthor[name=R1, color=red]{R1}
\definechangesauthor[name=R2, color=blue]{R2}

\begin{document}

\begin{frontmatter}

\title{A taxonomy of grain boundary migration mechanisms via displacement texture characterization}

\author[1,3]{Ian Chesser\corref{cor1}}
\ead{ichesser@gmu.edu}
\author[2]{Brandon Runnels}
\ead{brunnels@uccs.edu}
\author[1]{Elizabeth Holm}
\ead{eaholm@andrew.cmu.edu}

\cortext[cor1]{Corresponding author}
\address[1]{Department of Materials Science and Engineering, Carnegie Mellon University, Pittsburgh, PA, United States}
\address[2]{Department of Mechanical and Aerospace Engineering, University of Colorado, Colorado Springs, CO, United States}
\address[3]{Department of Physics and Astronomy, George Mason University, Fairfax, VA, United States}

\cortext[cor]{I. Chesser, Department of Physics and Astronomy, George Mason University, 4400 University Drive, Fairfax, VA, USA.}

\begin{abstract}

Atomistic simulations provide the most detailed picture of grain boundary (GB) migration currently available. Nevertheless, extracting unit mechanisms from atomistic simulation data is difficult because of the zoo of competing, geometrically complex 3D atomic rearrangement processes. In this work, we introduce the displacement texture characterization framework for analyzing atomic rearrangement events during GB migration, combining ideas from slip vector analysis, bicrystallography and optimal transportation. Two types of decompositions of displacement data are described: the shear-shuffle and min-shuffle decomposition. The former is used to extract shuffling patterns from shear coupled migration trajectories and the latter is used to analyze temperature dependent shuffling mechanisms. As an application of the displacement texture framework, we characterize the GB geometry dependence of shuffling mechanisms for a crystallographically diverse set of mobile GBs in FCC Ni bicrystals. Two scientific contributions from this analysis include 1) an explanation of the boundary plane dependence of shuffling patterns via metastable GB geometry and 2) a taxonomy of multimodal constrained GB migration mechanisms which may include multiple competing shuffling patterns, period doubling effects, distinct sliding and shear coupling events, and GB self diffusion.  

\end{abstract}

\end{frontmatter}


\section{Introduction}
\label{sec:intro}

Grain boundaries (GBs) are mobile planar defects in polycrystalline materials with structures and motion mechanisms that comprise a large phase space. Controlling GB structure and mobility is an important challenge during the processing of metallic and ceramic materials \cite{petch1953cleavage,chiba1994relation,randle2010grain,sangid2013physics,chookajorn2012design,shimada2002optimization,rollettrecrystallization,zhang2013stress,rupert2009experimental,li2013incoherent,jin2014annealing,lqcke1992texture}. Significant progress has been made in GB migration science over the last several decades at the atomic scale via a combination of atomistic simulations and experiments. Atomistic simulations, augmented by increasingly powerful computing resources and data analysis tools, have enabled high-throughput surveys of grain boundary properties and the discovery of new grain boundary structures and migration mechanisms \cite{olmsted2009survey,mishin2010atomistic,meiners2020observations}. In-situ imaging of grain boundary migration at the atomic scale has confirmed several theories that, until recently, were only supported by molecular dynamics (MD) simulations. For example, simultaneous grain rotation and shrinkage was recently observed in a 2-D system, confirming a main prediction of the Cahn-Taylor model \cite{ren2020grain,cahn2004unified}. Disconnection nucleation and growth has been observed directly during shear driven motion of grain boundaries in bicrystals \cite{wei2021direct}. Despite these synergistic findings, experiments raise hard questions about the impact of constraints and coupled processes on GB migration in real microstructures. Low dimensional structural phase transformations at GBs can dramatically impact grain boundary mobility \cite{meiners2020observations,krause2019review,Cantwell2014}. Large scale grain growth experiments show that the same crystallographically defined GB can move at very different speeds depending on local microstructure \cite{zhang2018three,zhang2020grain}. The impact of metastable GB structure and constraints on GB migration is still not well understood at the atomic scale. 

There have been a variety of post-processing techniques applied to MD simulations to characterize atomic displacement patterns in space and time during GB migration. Spatial displacement analysis methods include slip vector analysis \cite{coleman2014effect,priedeman2017role,bair2019antithermal}, microrotation analysis and other computations inspired from continuum mechanics \cite{tucker2011continuum,tucker2012investigating}. These methods have been particularly useful in analyzing ordered shuffling patterns of low $\Sigma$ GBs, where it has been concluded that ordered shuffling patterns are a necessary but insufficient condition for fast, non-thermally activated grain boundary migration \cite{bair2019antithermal}. Such ordered atomic rearrangement processes resemble those occurring during twinning and martensitic phase transformations \cite{hirth2016disconnections,therrien2020minimization}. Although these spatial displacement analysis methods are agnostic to unit mechanism, they are difficult to interpret in noisy settings where multiple mechanisms may compete at high temperatures or stresses. 

Another class of \textit{spatiotemporal} methods from statistical physics enables analysis of characteristic length and time scales associated with GB migration. The Van Hove correlation function and self intermediate scattering function, when computed over time intervals spanning atomic vibrations and atomic diffusion, are useful measures of dynamic heterogeneity and structural relaxation occurring during GB migration \cite{zhang2006characterization,zhang2006simulation,zhang2009grain,zhang2007atomic}. On the basis of these techniques, it has been argued previously that atomic rearrangements during GB migration bear strong resemblance to dynamic heterogeneity in supercooled glass forming liquids \cite{zhang2006characterization,zhang2006simulation}. For instance, quasi-1D clusters of mobile atoms called ``strings'' were observed during migration that increase in average length with decreasing temperature, similar to supercooled liquids \cite{zhang2009grain}. Stringlike displacements have also been observed during GB self diffusion \cite{mishin2015atomistic}. Although it is is well known that GB diffusion and migration can coexist, it is unclear how different types of atomic rearrangements such as strings and collective shuffles limit the rate of boundary migration under different conditions including varying interface geometry, temperature and driving force. Although spatiotemporal analysis methods are very useful for analyzing GB migration, they require migration trajectories with high time resolution that are expensive to analyze for a large parameter space. 

In this work, we aim to infer grain boundary migration mechanisms from two snapshots of grain boundary motion before and after motion. This set-up is similar to comparing TEM images before and after migration, except that in MD simulations the identities of distinct atoms are known. We approximate GB migration trajectories as probability distributions over shuffling vector lengths and orientations in the dichromatic pattern. We call these distributions \textit{displacement textures}, a term inspired by spin textures in magnetic materials. The displacement texture characterization procedure for analyzing atomic shuffles during GB migration can be seen as a 3-D generalization of slip vector analysis. Unlike standard slip vector analysis, however, displacement texture characterization accounts for multiple reference frames associated with symmetries in the dichromatic pattern, similar to the topological model of Hirth and Pond \cite{hirth2016disconnections,hirth2019topological}. Our framework allows inference of shuffling patterns for distinct disconnection modes. Disconnections are unit defects responsible for shear coupled migration and possess both sliding (burgers vector) and rotational (step) character. Multiple disconnection types with closely spaced nucleation and migration energy barriers may compete at finite temperature, leading to interesting phenomena such as GB roughening (a structural transition which leads to the onset of significant motion at finite temperature) and zig-zag mode switching under constraints \cite{thomas2017reconciling}. It is shown in this work that displacement texture analysis allows inference of distinct disconnection types from multi-modal MD trajectory data where traditional indicators of disconnection type such as coupling factor are difficult to interpret.

In order to construct a taxonomy of GB migration mechanisms via displacement texture analysis, we consider a diverse set of migration scenarios including the motion of twin boundaries, disconnection mediated shear coupled motion, and constrained mixed mechanism scenarios involving shuffling, sliding and GB self diffusion.  In combination with displacement texture characterization, a main analysis method of this work is a recently introduced forward model for GB migration which enables the interpretation of noisy and multimodal displacement data \cite{OT}. The main idea of the forward model is to generate a large catalog of GB migration mechanisms to compare to MD data. The forward model is informed by bicrystallography and allows enumeration of spatial displacement patterns for different misorientations, microscopic translations and disconnection modes. The forward model is probabilistic, enabling the prediction of competing transformations in the dichromatic pattern with a heuristic temperature-like regularization parameter. It is shown in this work that the forward model greatly aids the interpretation of displacement textures from MD data. 

The displacement texture method is formally described in the first half of this paper with examples given in the second half. Two types of decompositions of displacement data are described in Section \ref{sec:dt}: the shear-shuffle and min-shuffle decomposition. The former is used to extract shuffling patterns from shear coupled migration trajectories and the latter is used to analyze temperature dependent shuffling mechanisms. As an application of the displacement texture framework, we characterize the GB geometry dependence of shuffling mechanisms for a crystallographically diverse set of mobile GBs in FCC Ni bicrystals. Four examples of scientific interest include analysis of the boundary plane dependence of shuffling for $\Sigma 3$ GBs, analysis of variation of shuffling patterns with misorientation, analysis of mixed sliding and migration in a $\Sigma 11$ twist GB, and analysis of roughened, multimodal migration of symmetric tilt GBs.

\section{Molecular dynamics methodology} 
\label{sec:MD}

We use the FCC Ni bicrystal dataset generated by Olmsted et. al in \cite{olmsted2009survey,olmsted2009survey2} as a starting point for MD simulations. To illustrate how migration mechanisms vary with crystal structure, we also use several bicrystal structures from the BCC Fe dataset generated by Ratanaphan et. al \cite{ratanaphan2015grain}. These datasets span a variety of misorientations and boundary plane inclinations and are used to explore the impact of GB geometry on migration mechanisms. Boundary conditions for bicrystal structures were chosen to be periodic in the plane of the GB and free normal to the GB plane with 1 nm thick rigid slabs defined at each free surface. All bicrystals were resized to have dimensions larger than 16 nm in the interface plane normal direction ($\hat{x}$ direction) with cross sections larger than $3$ nm $\times$ $3$ nm. All GB structures were previously optimized via a grid search over microscopic translations $\bm{p}$ such that free energy is minimized at 0 K \cite{olmsted2009survey}. FCC Ni structures employ the Foiles-Hoyt EAM Ni potential \cite{foiles2006computation} as a model for interatomic bonding. The melting point of this potential is 1565 K. BCC Fe structures employ the Fe EAM Mendeleev potential \cite{mendelev2003development}. The melting point of this potential is 1772 K. Molecular dynamics simulations were performed in LAMMPS \cite{plimpton1995fast} and visualization performed in OVITO \cite{stukowski2009}. 


The ramped synthetic driving force method (rSDF) \cite{deng2017size} is used to generate grain boundary motion in bicrystal structures under both free and fixed boundary conditions. When fixed boundary conditions are enforced, a slab at the right end of the bicrystal is held in place. Before application of a driving force, relaxed bicrystal structures are equilibrated in a multistep procedure at a given set temperature. First, 0 K bicrystal structures are uniformly expanded by a pre-computed thermal expansion coefficient. Next, they are equilibrated in the NVT ensemble with free boundary conditions for 0.1 ns. The slab at the right end of the bicrystal is allowed to float in the plane normal direction to relieve stress in that direction. At the end of this procedure, normal stresses in the bulk grains are verified to fluctuate around zero with small amplitudes. After the equilibration run, boundary conditions are imposed (free or fixed) and a synthetic driving force is applied as a potential energy jump across the GB. The synthetic driving force is ramped upward from 0 at a rate of 0.2 meV/ps \cite{yusurvey} with the energy jump applied symmetrically across the GB. Each simulation run lasted 0.2 ns or was ended early if the GB had moved more than 4 nm. rSDF simulations for FCC Ni GBs were performed in the low to intermediate temperature range 100-1000 K (0.06-0.65 $T_m$). This temperature range was selected for ease of analysis of unit mechanisms. 

Depending on GB geometry, critical driving force values were found to vary from experimentally accessible driving force magnitudes (less than 10 MPa) for highly mobile GBs to high values (on the order of GPa) for low mobility GBs \cite{chesserout,yu2019survey,yusurvey}. Critical driving force values generally decrease with temperature, as described in \cite{yu2019survey}. Values of critical driving forces for free boundary conditions were previously tabulated in \cite{yusurvey}. The main goal of this work is analysis of migration mechanisms.

The ramped driving force is useful for stimulating changes in migration mechanism during a single simulation run. In particular, GBs that would ordinarily stagnate under constrained boundary conditions at constant driving force are often observed to continue migrating as driving force is increased via a diverse array of mode switching mechanisms. Our taxonomy of mode switching behavior extends the observations of zig-zag disconnection mode switching by Spencer and Srolovitz to a larger class of GBs \cite{thomas2017reconciling}. 

The Janssens implementation \cite{Janssens2006} of the synthetic driving force was used in this work because of its computational efficiency. An individual 0.2 ns run of the rSDF method took several hours on a 64 cpu machine compared to the original ECO method \cite{ulomek2015enfJergy} which took several days per run. The Janssens driving force has been shown to give consistent migration mechanism results compared to other techniques, including strain driven motion \cite{coleman2014effect}, random walk motion \cite{Trautt2006} and the ECO implementation of the synthetic driving force \cite{ulomek2015enfJergy}. Nevertheless, a deficiency of the Janssens implementation is that nearest neighbor rearrangement events of atoms in the GB core can lead to discontinuous changes in the applied synthetic energy which violate conservation of energy in the simulation. Mobilities computed with the ECO and Janssens method have shown similar trends with temperature and driving force but have systematic differences in magnitude with Janssens mobility values being lower than ECO values. At the time of writing, a more computationally efficient version of the ECO driving force has been published with significantly faster force calculations \cite{schratt2020} compared to the original version. Consistency of the mechanism results from this work with the new ECO method should be checked in future work.

\section{Displacement Texture}
\label{sec:dt}

\subsection{Informal statement of main ideas}\label{sec:main_ideas}

\begin{figure}[th]
    \centering\leavevmode
    \includegraphics[width=0.9\textwidth]{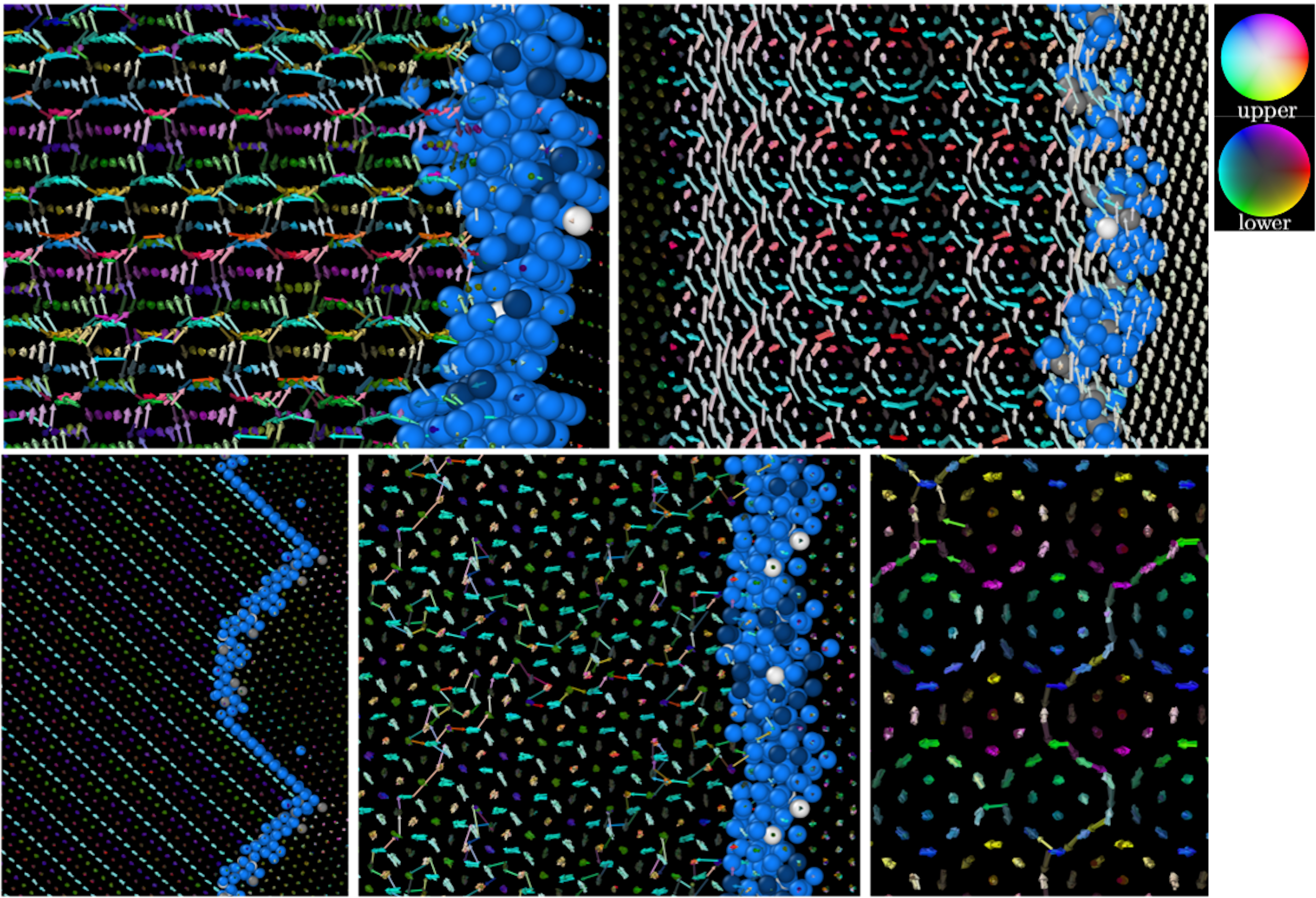}
    \caption{These images show diverse displacement textures swept out by mobile GBs considered in this work. Atoms in the grain boundary core are shown explicitly, colored by non-FCC coordination (HCP = light blue, BCC = grey, SC = white, OTHER = dark blue) in OVITO \cite{larsen2016}. Atomic displacements are colored by orientation in the sample frame of each bicrystal using the default ipfHSV color scheme in MTEX \cite{MTEX}. The scale of the top left image is 5 nm x 3 nm. Top left: $\Sigma 45$ \hkl<210> 84$^\circ$ tilt, BCC Fe (id 102), 100 K. Top right: $\Sigma 37$ \hkl<100> 18.9$^\circ$ tilt, FCC Ni (id 216), 200 K. Bottom left: $\Sigma 3$ mixed tilt twist, FCC Ni (id 337), 500 K. Bottom middle: $\Sigma 5$ \hkl(430) asymmetric tilt, BCC Fe (id 36), 500 K. Bottom right: $\Sigma 51$ \hkl(110) 16$^\circ$ twist GB, FCC Ni (id 356), 400 K. Boundary conditions are fixed in all of the simulation snapshots shown above. For full crystallographic information corresponding to each GB id, see the supplementary information of \cite{olmsted2009survey,ratanaphan2015grain}}
    \label{fig:intro}
\end{figure}

Colored displacement textures highlight strikingly ordered patterns of atomic motion during grain boundary migration (Figure \ref{fig:intro}). Displacement textures comprise the linear net displacement vectors obtained by subtracting the atomic positions after GB migration from those before GB migration. The goal of displacement texture analysis in this work is twofold: 1) infer and separate migration mechanisms within individual GB migration trajectories and 2) quantify how displacement patterns vary over a large parameter space including different GB types, temperatures, driving forces and variable boundary conditions. 

The displacement textures in Figure \ref{fig:intro}  illustrate a diverse range of migration mechanisms. Some GBs, such as the  $\Sigma45$ $\hkl<210>$ $84^\circ$ symmetric tilt GB in BCC Fe (top left), exhibit visually striking helical displacement patterns that tile space in 3D. The visually apparent order of these displacement patterns stands in stark contrast to the disordered core structure of the GB. 
Other GBs, such as faceted $\Sigma 3$ mixed tilt-twist GBs (lower left), exhibit ordered layers of displacements. Many tilt and twist GBs sweep out \textit{vortex} displacement patterns which circulate around fixed points of near zero displacement (upper and lower right). In some cases, such as migration of the $\Sigma 5$ $\hkl(430)$ asymmetric tilt (lower middle) and $\Sigma 51$ \hkl(110) 16$^\circ$ twist GB (lower right), networks of displacements are apparent that are distinct from the overall ordered displacement pattern. These mixed displacement types indicate the coexistence of GB migration and GB self diffusion. Nonuniform displacement textures are indicative of a change in migration mechanism or a mixture of migration mechanisms and are common during migration simulations with variable driving force \cite{yusurvey}. An example of nonuniform migration is the constrained, driven migration of a  $\Sigma 37$ \hkl<100> 18.9$^\circ$ symmetric tilt GB (top right) which shows shear deformation during an early part of its motion and vortex displacement patterns with less apparent shear later in its motion. Interpretation of non-uniform migration trajectories is a characterization challenge which requires defining and separating unit mechanisms.  

To allow for quantitative analysis of displacement patterns during GB migration, we will formally define displacement texture as a probability distribution over shuffling lengths and directions in the dichromatic pattern in Section \ref{sec:dtxdef}. In Sections \ref{sec:tax} and \ref{sec:ss}, we give prerequisite background information about the dichromatic pattern and the shear-shuffle decomposition. The shear-shuffle decomposition allows identification of shuffle displacements corresponding to arbitrary disconnection modes. In Section \ref{sec:minshuff}, we review key aspects of a previously developed forward model for shuffle patterns \cite{OT} to set the stage for interpreting the impact of geometric degrees of freedom (microscopic shift, misorientation and boundary plane inclination) on migration mechanisms. Details of practical implementation of displacement texture analysis are given in \ref{sec:imp1}. 

\begin{figure}[th]
    \centering\leavevmode
    \includegraphics[width=0.9\textwidth]{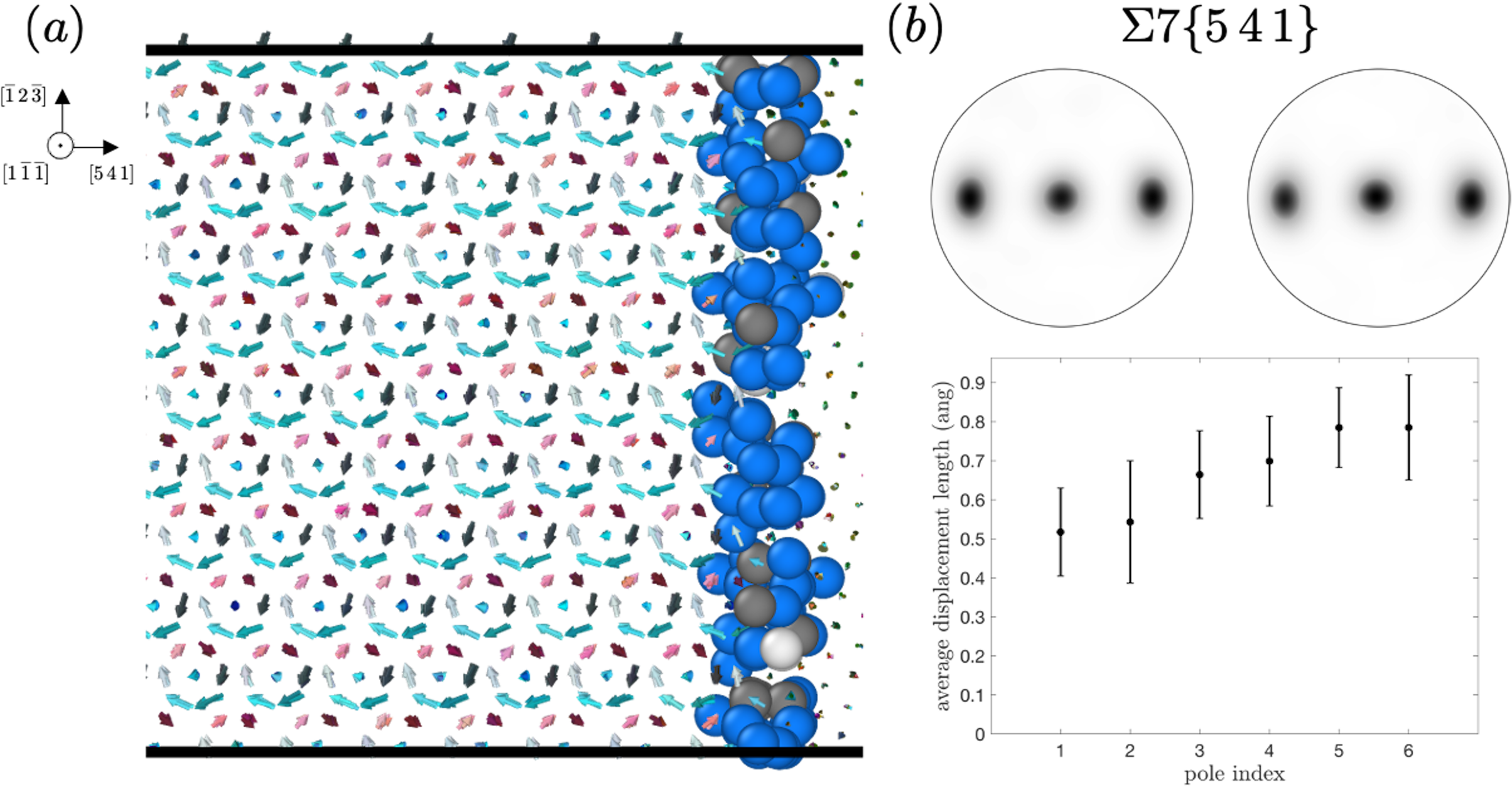}
    \caption{(a) Example of a displacement texture for a $\Sigma 7$ GB (id 26) at 300 K under constrained boundary conditions. The hexagonal shuffling pattern highlights the \hkl(111) symmetry of the dichromatic pattern for this GB. Atoms at fixed points at the center of each hexagon do not need to move to accommodate reorientation and correspond to coincident cites in the dichromatic pattern (b) The displacement texture is represented by stereographic projections of the displacements in the upper and lower hemisphere (upper panel). In this case, the six non-zero shuffling vectors are equally probable but have different displacement lengths (lower panel), all of which are shorter than the nearest neighbor distance $\sigma = 2.49 $ angstroms. Error bars are computed as the standard deviation in displacement length for the set of atoms with a given shuffle orientation. Asymmetric shuffle lengths result from a preferred microscopic translation in the dichromatic pattern. }
    \label{fig:dtx_example}
\end{figure}

The main graphical representation of displacement texture in this work is an inverse pole figure which we call the \textit{displacement pole figure} (Figure \ref{fig:dtx_example}). In the displacement pole figure, orientations of displacement vectors swept out during GB migration are plotted in stereographic projection in the sample frame of the bicrystal. Further details of the practical implementation of displacement texture analysis are given in \ref{sec:imp1}. 

\subsection{The coherent and translated dichromatic pattern}\label{sec:tax}

Atomic displacements swept out during conservative grain boundary migration connect sub-lattices of the dichromatic pattern. In this section, we formally define two types of dichromatic patterns useful for analyzing GB migration, borrowing terminology from the topological model of Hirth and Pond \cite{pond2019topological,hirth1996steps,pond1983bicrystallography}. 

 Let $\mathcal{X},\mathcal{Y}\subset\mathbb{R}^3$ be the set of all atomic positions in the lattice 1 and lattice 2 configurations of a bicrystal. We choose the reference atomic positions $\mathcal{X},\mathcal{Y}$ such that $|(\mathcal{X}\cap\mathcal{Y})|\ge1$; in other words, there is at least one common (``coincident'') site 
(Equality holds in the case of lattices rotated by an irrational angle, since the only coincident point is at the origin \cite{runnels2017projection}).
 $\mathcal{X}\cap\mathcal{Y}$ is the Coincident-Site Lattice (CSL) for the bicrystal and the union $\mathcal{X}\cup\mathcal{Y}$ is the Coherent Dichromatic Pattern (CDP). The CDP is the most commonly depicted dichromatic pattern in the literature, but is not the reference frame that is most relevant for grain boundary migration.
 
In real or simulated boundaries, there is also a rigid microscopic shift ($\bm{p}\in\mathbb{R}^3$) between the two sub-lattices in the dichromatic pattern, which is conventionally understood to minimize the grain boundary energy for a stationary boundary. The actual atomic positions in lattice 2 are denoted $x=\mathcal{Y}+\bm{p}$. 
\footnote{The use of lowercase $x$ is following traditional continuum mechanics parlance that denotes undeformed and deformed positions by uppercase and lowercase letters, respectively.}
It is useful to construct a modification of the CDP: $\mathcal{X}\cup x$, which is referred to as the Translated Dichromatic Pattern (TDP). The raw atomic displacements swept out by GB migration during MD simulations are represented in the TDP. Displacements in the TDP may be converted to the CDP by subtraction of the microscopic shift vector $\bm{p}$. The CDP provides a convenient reference frame for analyzing the crystallography of shuffling directions (see results for $\Sigma 3$ GBs in Section \ref{sec:sig3micro}). However, since $\bm{p}$ is often unknown in practice (for GBs in a polycrystal), displacements in the TDP are most directly accessible to simulations and experiments. 

 \begin{figure}[th]
    \centering\leavevmode
    \includegraphics[width=1.0\textwidth]{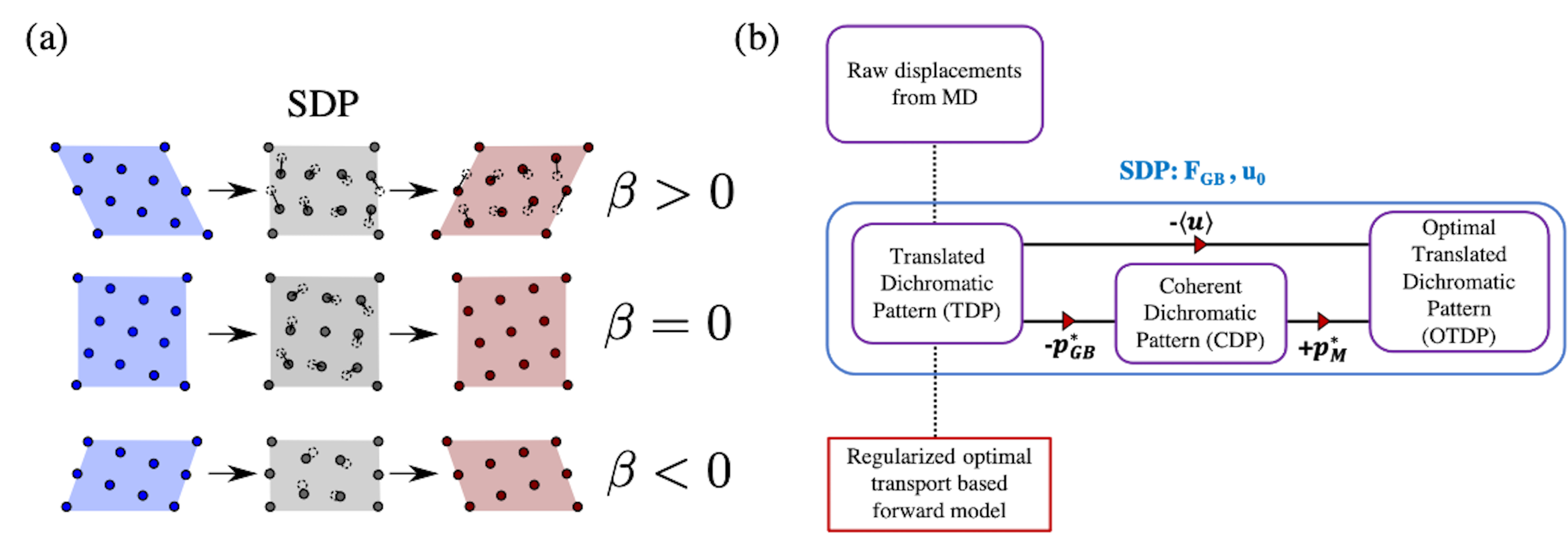}
    \caption{(a) The shear shuffle decomposition is shown for three different types of transformations in the dichromatic pattern for the same boundary: a positive coupling mode ($\beta > 0$), a zero shear transformation ($\beta = 0$) and a negative coupling mode ($\beta < 0$). The net deformation $\bm{F^{GB}}$ may be decomposed into a combination of shuffle displacements (middle frame) and shear displacements (arrows). The shuffling dichromatic pattern (SDP) is associated with the zero shear transition state for a given deformation. (b) provides a summary of dichromatic pattern types for a fixed choice of shear deformation $\bm{F^{GB}}$ and normal coordinate $\bm{u_0}$. These three types of dichromatic patterns are distinguished by three rigid translations in the SDP. The coherent dichromatic pattern (CDP) is associated with zero translation. The translated dichromatic pattern (TDP) is obtained from the CDP by a microscopic shift $\bm{p^*_{GB}}$ typically associated with energy minimization. The optimal translated dichromatic pattern (OTDP) is associated with zero net displacement per atom and minimal net shuffling distance and can be obtained from the TDP by subtracting a translation $\langle \bm{u} \rangle$ or from the CDP by adding a translation $\bm{p^*_M}$.}
    \label{fig:workflow}
\end{figure}


\subsection{The shear shuffle decomposition}\label{sec:ss}

The notion of a shear shuffle decomposition goes back to early work on twinning by Bilby and Crocker \cite{bilby1965theory,bevis1969twinning,christian1995deformation}. It was recognized that the atomic positions of two grains across a twin interface cannot always be matched by a pure shear operation. In other words, depending on the twin orientation relationship, shuffles are required in addition to shear displacements to accommodate crystal rotation during twin formation (Figure \ref{fig:workflow}a). In cases where normal motion is coupled to sliding during GB migration, the raw displacements swept out behind the moving boundary can be decomposed into shuffle and shear displacements arising from the normal (step) and sliding (burgers vector) character of disconnection motion. In this section, we describe the process of identifying shear and shuffle displacements directly from MD data for arbitrary disconnection types. 

Given a grain boundary shear deformation gradient $\bm{F}^{GB}\in GL(3)$ (swept by reorientation of grain 1 to grain 2), general atomic displacements $\bm{x}_i$ are represented in the TDP as:
\begin{align}
  \bm{x}_i = \bm{F}^{GB}\bm{X}_i + \bm{p} + \bm{u}_i
  \ \ \ 
  \bm{x}_i,\bm{X}_i \in x \times \mathcal{X},
\end{align}
where $\bm{u}_n$ is a local atomic shuffle and $\bm{p}$ is the microscopic shift vector chosen on the basis of minimum GB energy. The grain boundary shear deformation gradient $\bm{F}^{GB}$ usually has a single nontrivial component in the sample frame, the scalar coupling factor ($\beta_n = \frac{b_n}{h_n}$), which depends on disconnection geometry $(b_n,h_n)$ for mode $n$.

Atomic displacements are said to be Cauchy Borne (C-B) if the shuffling vector $\bm{u}_i = \bm{0}$ and all atomic displacements point in the direction of the net shear displacement. In the parlance of the shear shuffle decomposition, C-B displacements are \textit{shears}. Apart from a few special cases, such as shear coupled motion of the coherent twin and other tilt boundaries \cite{mompiou2010smig}, GB migration (and disconnection motion) are not C-B but instead consist of both shears and shuffles. In fact, many GBs in this survey, including most $\Sigma 3-11$ GBs, move with zero net shear deformation ($\bm{F}^{GB}=\bm{I}$) such that shuffles dominate the reorientation geometry. It should be cautioned that shears and shuffles are defined relative to the operative disconnection mode or sliding mechanism. The same spatially defined atomic displacements may participate in different disconnection modes and may therefore be classified as shears or shuffles depending on context. It is also possible for a sequence of alternating shear events to average to zero net deformation, as has been observed during the zig-zag GB migration of a \hkl<111> tilt GB \cite{thomas2017reconciling,olmsted2007grain,holm2010grain,chen2020grain}. 

 \begin{figure}[th]
    \centering\leavevmode
    \includegraphics[width=0.8\textwidth]{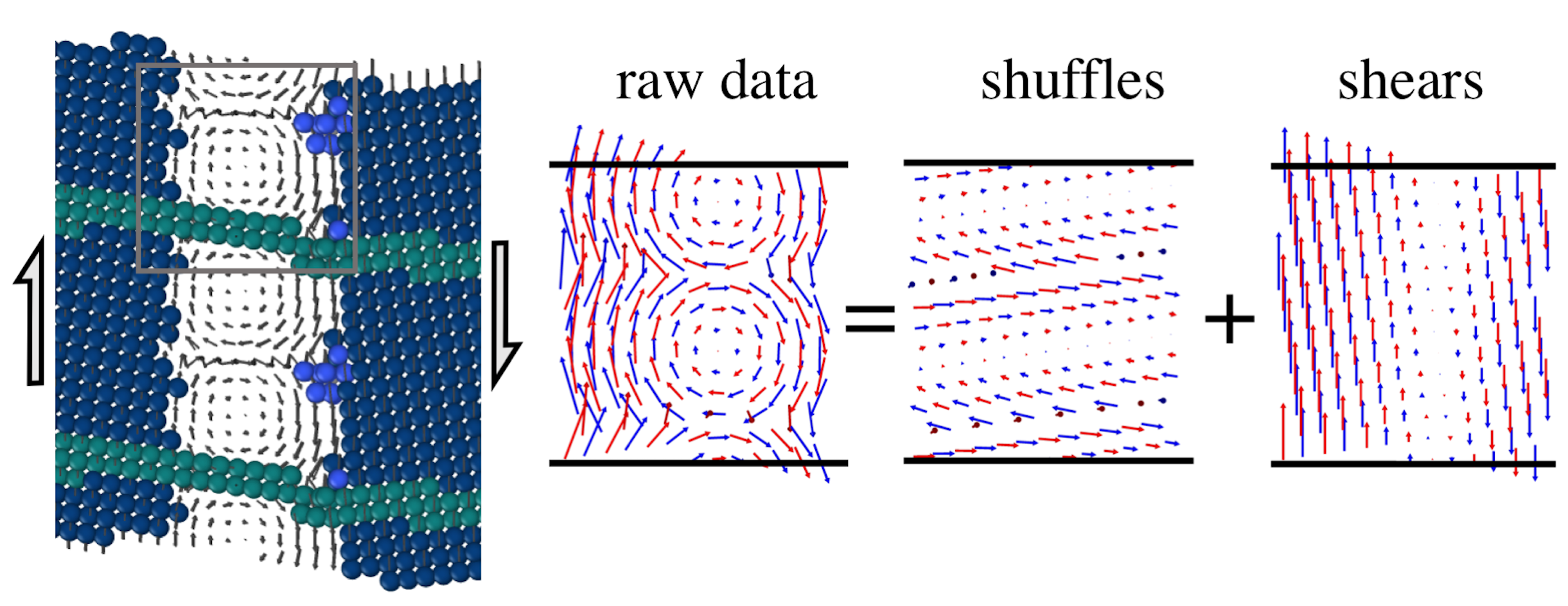}
    \caption{Example of the shear shuffle decomposition for coupled migration of the $\Sigma 25$ \hkl<100> tilt $16.3^\circ$ GB with an assumed \hkl(100) type coupling mode. Raw MD data from an energy jump driven migration simulation at 100 K is shown on the left with lateral fiducial markers shown in blue-green and atoms swept out by the moving GB deleted. In the decomposed data, displacement vectors are colored by position into the page along the tilt axis  \cite{cahn2006coupling,chesserout}.}
    \label{fig:ex_ss}
\end{figure}

The Shuffling Dichromatic Pattern (SDP) is the appropriate dichromatic pattern representation for analyzing shuffling patterns from MD data. This representation was recently introduced by Hirth in the context of twin formation in minerals \cite{hirth2019topological}. Construction of the SDP requires a shear deformation $\bm{F}^{GB}$ related to a known coupling factor to be measured and subtracted from the displacement pattern in the TDP.  The SDP representation is not unique, and a set of SDP representations can be indexed by disconnection mode or coupling factor. Note that, since displacement texture is defined as a distribution of shuffling vectors, it is necessary to remove the affine shear deformation from the raw displacement field obtained in MD simulations.  

There are several choices for how to subtract shear from the raw displacement pattern in order to recover shuffling vectors, including subtracting shear from one or both sublattices in the TDP. In Figure \ref{fig:workflow}a, the SDP is constructed by symmetrically subtracting shear deformations with components $\beta_i/2$ and $-\beta_i/2$ from each of the sublattices of the TDP. In Figure \ref{fig:ex_ss}, the shear shuffle decomposition is applied to MD data for a $\hkl<100>$ $16.3^\circ$ tilt GB which exhibits shear coupling with coupling factor $\beta = 0.28$. This coupling factor corresponds to the well known \hkl(100) type coupling mode \cite{cahn2006coupling,chesserout}. The raw displacements swept out by shear coupled motion form a vortex pattern with increasing shear displacements away from the vortex centers. To recover the shuffling pattern associated with this coupling mode, shear is symmetrically subtracted about the vortex centers at a position $\bm{u_0}$ normal to the GB. This subtraction recovers a layered shuffling pattern (middle panel of Figure \ref{fig:ex_ss}) associated with migration of this GB in the \hkl(100) coupling mode for a specific microscopic shift $\bm{p}$. We note that $\bm{u_0}$ defines a plane about which no deformation is subtracted and that disconnection step height $h$ can be inferred from the periodicity of the SDP with respect to $u_0$ \cite{OT}. A systematic numerical algorithm for calculating $\bm{F}^{GB}$ and $u_0$ from bicrystal coordinates before and after migration is described in \ref{sec:appendix_CB_deformation}. 

\subsection{Optimal transportation and the min-shuffle decomposition}\label{sec:minshuff}

\begin{figure}[th]
    \centering\leavevmode
    \includegraphics[width=0.8\textwidth]{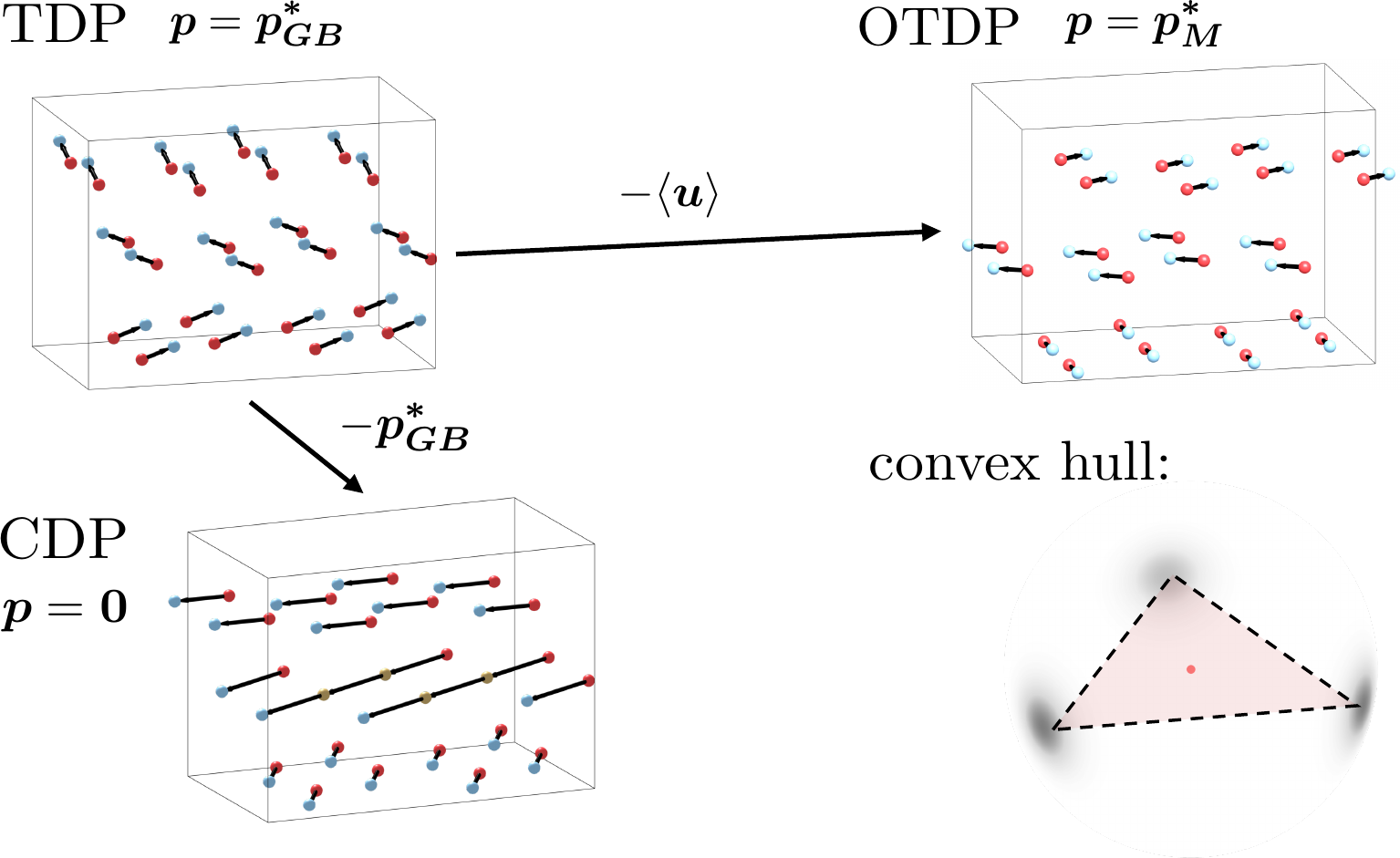}
    \caption{Min-shuffle transformation prediction for the $\Sigma 3$ \hkl(110) GB with zero coupling factor $\bm{F}^{GB} = \bm{I}$. Shuffles are determined via the optimal transport based forward model at low $\epsilon$ in the TDP with a microscopic shift vector $\bm{p} = \bm{p^*_{GB}} $chosen to match the energy minimizing shift in the relaxed bicrystal structure at 0 K. Shuffling patterns in the CDP and OTDP are obtained by distinct rigid translations $\bm{p^*_{GB}}$ and $\U$ of one sublattice in the TDP. The OTDP is distinguished by zero net displacement and shuffles with minimal total length for this 3-shuffle transformation. The convex hull of shuffle orientations in the OTDP is an equilateral triangle.}
    \label{fig:refexample}
\end{figure}

In this section, we describe a decomposition of shuffling vector data (in the SDP) which has practical and theoretical significance for understanding the impact of GB geometry on shuffling patterns. It has been hypothesized by Hirth et. al \cite{hirth2016disconnections} that the most likely shuffling pattern during GB migration is the one that minimizes total shuffle distance in the dichromatic pattern. This hypothesis was used as a starting point to develop a forward model for GB migration based on optimal transportation in the dichromatic pattern \cite{OT}. The mathematical framework for this forward model is described in \cite{OT}. An important concept in this modeling framework is that shuffling patterns can be locally optimized over microscopic shifts in the dichromatic pattern to minimize total shuffling distance. We denote the \textit{min-shuffle} transformation as the shuffling pattern with minimum total shuffling distance for a given disconnection mode. There exists a microscopic shift $\bm{p} = \bm{p^*_M}$ associated with the min-shuffle transformation which defines the Optimal Translated Dichromatic Pattern (OTDP). A summary of dichromatic pattern types is given in Figure \ref{fig:workflow}b along with example shuffling patterns for a $\Sigma 3$ GB in Figure \ref{fig:refexample}. The shuffling pattern in the OTDP has minimum net shuffle distance and can easily be obtained from shuffling vectors in the TDP by subtraction of the vector $\U$ computed as the mean shuffling vector in the TDP. $\U$ can be thought of as a sliding vector which transforms a metastable GB structure to another (possibly unstable) structure which permits shorter shuffles. 

Two important aspects of the min-shuffle transformation which will be discussed further in this work are degeneracy and regularization. In the degenerate case, multiple choices of $\bm{p}$ may correspond to the same locally minimum shuffling cost in the dichromatic pattern. In this situation, the min-shuffle transformation is not unique and different symmetric variants of shuffling patterns may exist. One physical manifestation of this degeneracy is given in Section \ref{sec:imp1} where period doubling effects are observed in the shuffling patterns of the $\Sigma 3$\hkl(112) GB. Related to degeneracy is the concept of regularization. The addition of regularization to the forward model mixes multiple distinct shuffling patterns with similar shuffling costs, allowing prediction of competing shuffling patterns in the dichromatic pattern. The regularization parameter $\epsilon$ is a heuristic temperature-like parameter which uniformly mixes all shuffling types in the limit $\epsilon \rightarrow \infty$. In \cite{OT}, competing shuffling patterns for the $\Sigma 5$ \hkl(100) twist GB were observed in MD simulations which have the same shuffle types as predicted by the regularized optimal transport model.

\subsection{Formal definition of displacement texture}\label{sec:dtxdef}

\begin{figure}
    \centering
    \includegraphics[width=\linewidth]{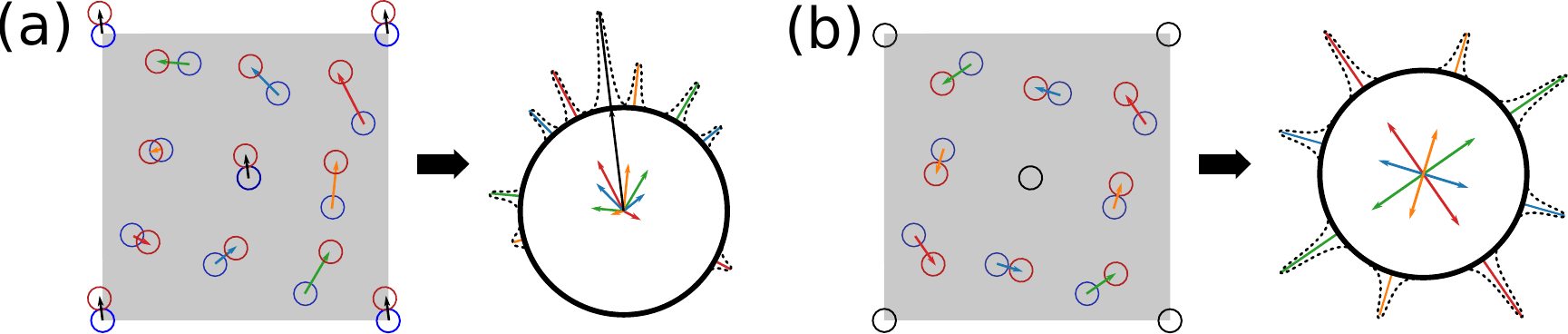}
    \caption{
        Example 2D displacement textures in the SDP depicting shuffles from (a) raw MD data (TDP) and (b) distance minimizing shuffles in the OTDP.
        Black circles correspond to atoms with no displacement; blue circles are atoms prior to displacement and red are atoms after displacement.
        The different colored arrows on the left illustrate the different fundamental shuffle directions.
        The right side of the figure superimposes all shuffles relocated to the origin.
        The displacement texture is plotted on the black circle as a polar plot.
        The dotted line corresponds to the mollified/smoothed texture corresponding to thermal fluctuations.
        Although (a) and (b) both correspond to the same type of shuffling pattern, the texture as plotted in (a) is artificially complex, whereas (b) distills the data into its fundamental peaks. Physically, (a) and (b) correspond to zero shear motion of the same GB with different choices of microscopic shift.}
    
    \label{fig:disptexexample2d}
\end{figure}

    A displacement texture is a probability distribution defined over a sphere, such that its magnitude at each point corresponds to the average amount of shuffle by atoms in that direction.
    Displacement textures enable the reduction of large quantities of atomistic data onto a compact domain, making it possible to identify statistically prevalent trends in atomic motion.
    Displacement textures can either be calculated from atomistic simulations or predicted using other models, and can account for discrete numbers of atoms or continuous distribution of atomic shuffles.
    Figure~\ref{fig:disptexexample2d} illustrates the concept of displacement textures for an artificial 2-D shuffling pattern. 

    In the remainder of this section, we present a formal definition of displacement texture in the language of measures.
    (While this is necessary in order to generalize to both the discrete and continuous cases, the reader may skip the remainder of Section~\ref{sec:dtxdef} without missing anything essential to the basic understanding of displacement textures.)
    We begin with the formal definition of displacement texture, which has the structure of a measure, defined as the following:

\begin{definition}
  The displacement texture measure $\mu$ is a measure on $S(2)$ for a set of shuffle displacements $\{\bm{u}_i\}$ such that
  \begin{align}
    \int_{\varsigma\subset S(2)}\bm{n}\,d\mu(\bm{n}) = \langle \bm{u} \rangle_\varsigma,
  \end{align}
  where $\langle \bm{u} \rangle_\varsigma$ is the average atomic displacement within $\varsigma$ and $\bm{n}$ is a unit normal vector.
\end{definition}

For instance, if $\varsigma=\{\bm{x}\in S_2:x_1\ge0\}$ (the positive $x_1$ hemisphere), the action of $\mu$ on $\varsigma$ produces the average atomic displacement in the $x_1$ direction.
In the specific case in which a discrete collection of shuffle displacements $\{\bm{u}_i\}$  is considered one may define $\mu$ as a sum of Dirac measures:
\begin{align}
  \mu(\bm{n}) = \frac{1}{N}\sum_i^N|\bm{u}_i| \, \delta\Big(\bm{n} - \frac{\bm{u}_i}{|\bm{u}_i|}\Big),
  \ \ \ N = \#(\{\bm{u}_i\}).
\end{align}
We now define the net shuffling vector $\U$,
\begin{align}\label{eq:convhull1}
  \U = \int_{B\subset S(2)}\bm{n}\,d\mu(\bm{n}).
\end{align}
In the special case where a finite number of discrete atoms are considered, it is possible to show that the above formula reduces to
\begin{equation}
   \U= \frac{1}{N}\sum_{i=1}^N\bm{u}_i,
\end{equation}
recovering simply the arithmetic mean of the shuffle displacements.

Recall that shuffle vectors in the OTDP are defined as

\begin{align}\label{eq:sdp}
\bm{u^*} = \bm{u} - \U
\end{align}

Shuffles in the OTDP satisfy $\langle \bm{u^*} \rangle = \bm{0}$. In practice, this constraint is convenient for symmetrizing displacement patterns and comparing them across a large parameter space irrespective of small perturbations. The shuffling vectors in the OTDP define a convex hull around the origin:
\begin{align}
  \langle \bm{u^*} \rangle = \bm{0} \in \operatorname{conv}\{\bm{u^*}_1,\ldots,\bm{u^*}_N\}.
\end{align}
This property is convenient for isolating and counting distinct shuffling vectors. 

The point measure definition of $\mu$ may be regularized using a mollifier (approximated by a truncated gaussian) $\psi_\varepsilon$ with variance $\varepsilon$, so that
\begin{align}
  (\mu\star\psi_\varepsilon)(\bm{n})  = t(\bm{n})\mathcal{H}^2,
\end{align}
with $\mathcal{H}^2$ the 2-Hausdorff measure, and 
\begin{align}
  t(\bm{n}) = \frac{1}{N}\sum_i^N|\bm{u}_i| \, \psi_\varepsilon\Big(\bm{n} - \frac{\bm{u}_i}{|\bm{u}_i|}\Big),
\end{align}
where $t(\bm{n})$ is refered to here and subsequently as the {\it thermalized displacement texture}. Thermalization can thus be applied synthetically by defining $\psi_\varepsilon$ as  $\varepsilon\propto k_BT$.
This mimics the effect of thermal fluctuations through smearing of the displacement poles.\footnote{We note that, since displacement texture is defined over a 2D domain, thermalization only reflects variance in the tangential component of the shuffle vectors, not the radial
component.}
Different types of noise can be fit to the MD data. 


\subsection{Displacement pole figures: implementation and conventions}\label{sec:imp1}

The thermalized displacement texture is estimated directly from MD data via probability density estimation. The information contained in the displacement texture can be separated into shuffle length and orientation distributions. In this work, we pursue a visualization method for shuffle orientation distributions that optionally includes information about shuffle length. Displacement pole figures plot displacement orientations in stereographic projection relative to the sample frame of the bicrystal. The grain boundary orientation in the inverse pole figures is always the same; displacement vectors lying in the GB plane (for a flat GB) plot along vertical lines in the stereographic projections of the upper and lower hemisphere.
Displacement pole figures are generated in MTEX with a spherical harmonic fit with the default de la Vallee-Poussin kernel \cite{MTEX}.  Displacement magnitude may be added to the spherical harmonic fit as an additional weighting factor for each atom. In this work, displacement pole figures plotted in the TDP and CDP contain displacement length information, while displacement pole figures plotted in the OTDP contain displacement orientation information only. The weighted displacement pole figures are convenient for visualizing large, low probability displacements which are commonly associated with multi-hop GB diffusion events.

Displacement pole figures provide a convenient starting point for counting and classifying shuffle types during low $\Sigma$ (typically $\Sigma < 13$) GB migration and give insight into nearly continuous shuffle orientation distributions during high $\Sigma$ GB migration. In cases where distinct displacement poles are well separated in orientation space ($> 10^\circ$), a probability threshold is defined to count the number of unique displacement types. In this work, we determine a unique shuffle type as one corresponding to a maximum pole probability in the OTDP greater than 0.005 with mean displacement length larger than typical atomic vibrations ($> 0.04$ nm). For high $\Sigma$ GBs or high temperature GB migration, displacement length information must be incorporated into counting routines since distinct shuffles may exist with closely spaced orientations or significantly overlapping orientation distributions. For most accurate shuffle identification, shuffles enumerated from the forward model can be directly compared to the MD data at varying degrees of regularization as in \cite{OT}. We recommend that comparisons in noisy settings be performed in the OTDP. 

\section{Applications of displacement texture}\label{sec:apps}

\subsection{Characterization of $\Sigma 3$ twin boundary migration in FCC Ni}\label{sec:sig3}

In this section, we show that the boundary plane inclination variation of shuffling patterns for $\Sigma 3$ GBs in MD simulations takes a simple form that can be rationalized by distance minimization in the dichromatic pattern with respect to microscopic translations. This connection between microscopic translational degrees of freedom and boundary plane dependent shuffling patterns is found to be generic to all low $\Sigma$ GBs in our MD datasets for FCC Ni. 

Twin GBs with a $\Sigma 3$ misorientation are common in FCC materials and exhibit anisotropic mobility behavior depending on boundary plane inclination \cite{Humberson2019,aditi_science}. The most common types of $\Sigma 3$ GBs in FCC metals are the $\hkl(111)$ Coherent Twin Boundary (CTB) and the $\hkl(112)$ Incoherent Twin Boundary (ITB). The acronym ITB refers to any facet that is not $\hkl(111)$ type. ITBs form as orthogonal or near orthogonal facets along CTBs. Flat CTBs are known to be extremely immobile, and are the basis for many GB engineering efforts because of their low grain boundary energy, high yield strength and high thermal stability \cite{shimada2002optimization,lu2009strengthening,randle2010grain}. On the other hand, ITBs are highly mobile and ITB facet motion is observed during diverse processes such as recrystallization, grain growth, radiation damage and stress driven coarsening \cite{li2013incoherent,bufford2014situ}. 

\subsubsection{The importance of microscopic shifts in determining shuffling for different plane inclinations}\label{sec:sig3micro}

\begin{figure}[th]
    \centering\leavevmode
    \includegraphics[width=0.9\textwidth]{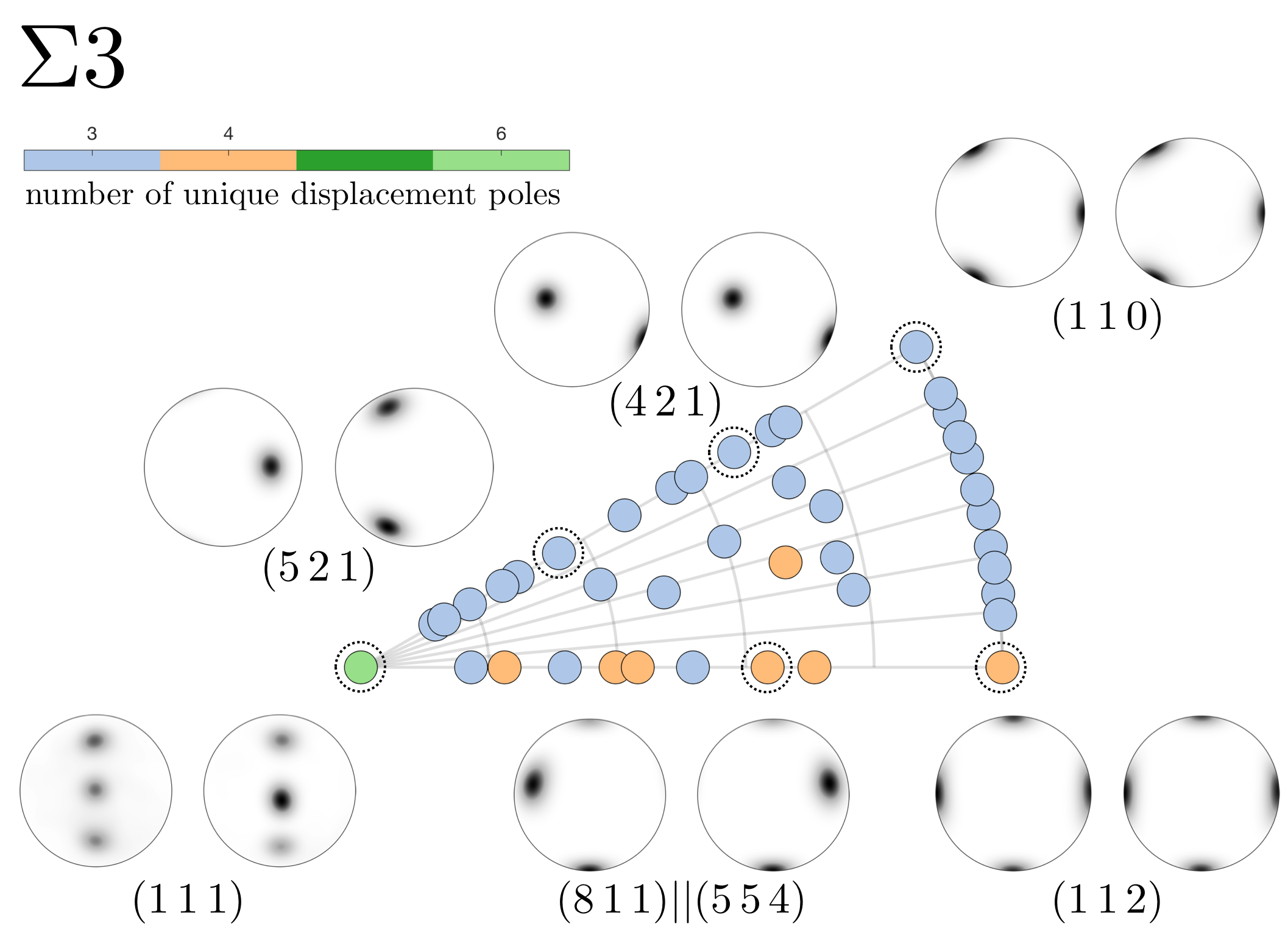}
    \caption{For $\Sigma 3$ GBs with varying GB plane inclination, multiple types of shuffling transformations are observed. Points in the boundary plane fundamental zone are colored by the number of unique shuffle vectors observed in the displacement texture of each GB at 300 K. Displacement pole figures in the OTDP illustrate three classes of transformations corresponding to \hkl{110} facet (3 shuffle), \hkl{112} facet (3-4 shuffle) or \hkl{111} facet (6 shuffle) controlled motion.}
    \label{fig:sig3inv}
\end{figure}

Only three distinct types of shuffling transformations are observed for 41 $\Sigma 3$ GBs in our dataset over a relatively large temperature range from 100-1000 K under both free and constrained boundary conditions. Most $\Sigma 3$ GBs move with zero net shear, with two exceptions (the \hkl(112) ITB and \hkl (111) CTB) discussed in Section \ref{sec:degen}. Displacement pole figures corresponding to migration are plotted in Figure \ref{fig:sig3inv}) for different boundary inclinations at 300 K. We define \textit{characteristic displacement textures} as shuffling patterns in the OTDP that are unique up to a rigid rotation. Characteristic displacement textures are observed with three, four and six unique displacement poles. 

The six pole transformation is only observed for the CTB. In contrast, the four pole transformation is observed for a sequence of $\hkl<110>$ asymmetric tilt GBs that move via glide of $\hkl(112)$ type ITB facets. The most common transformation in the MD data has only three poles and is observed during $\hkl(110)$ ITB facet glide. The rigid rotation of poles with GB geometry simply represents the spatial rotation of facets that control migration of these GBs. 

\begin{figure}[th]
    \centering\leavevmode
    \includegraphics[width=1.0\textwidth]{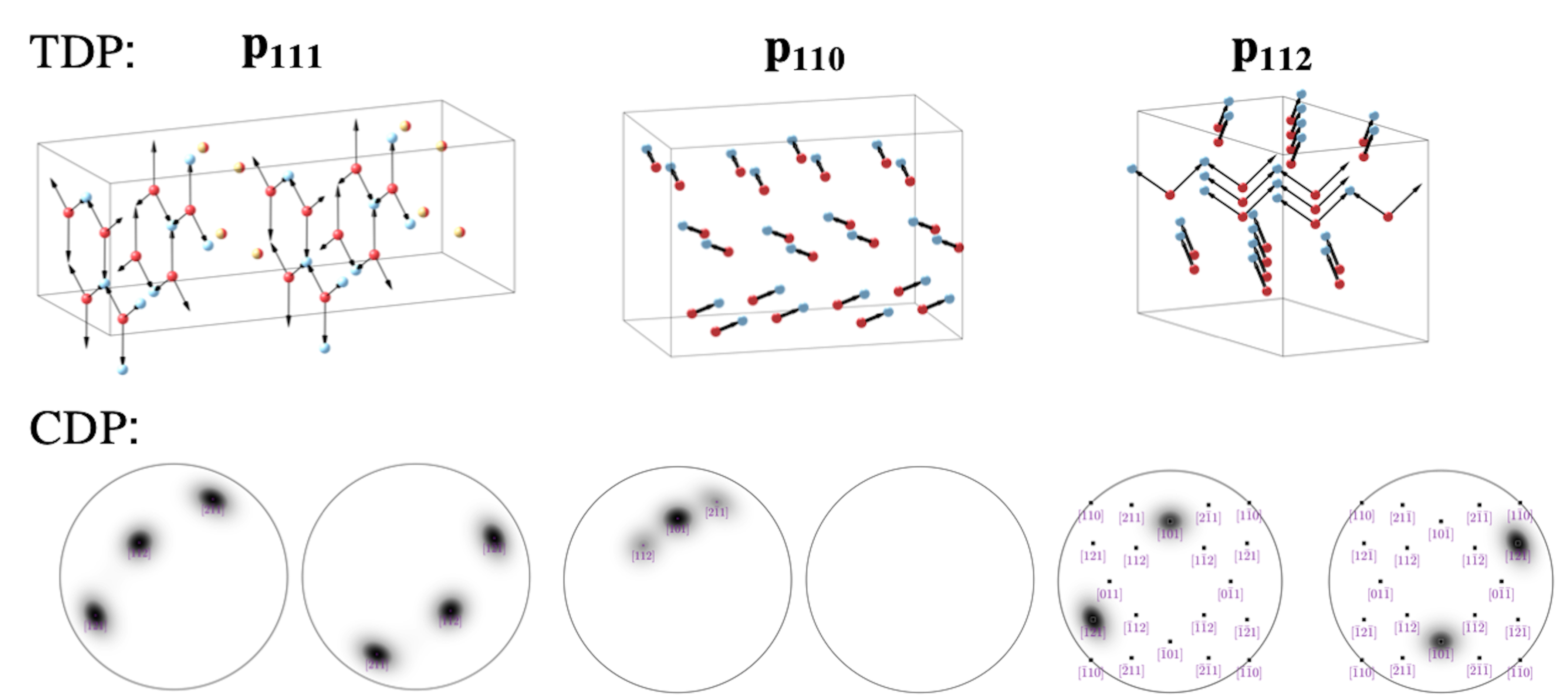}
    \caption{Min shuffle transformations for $\Sigma 3$ twin GBs in the TDP and corresponding displacement pole figures in the CDP. Degenerate displacements correspond to equal length displacements.}
    \label{fig:sig3pf}
\end{figure}

One target application of displacement texture analysis is the crystallographic identification of shuffling vectors in the crystal frame of each grain. Recall that the choice of microscopic shift $\bm{p}$ impacts shuffling directions. When shuffling directions are expressed in the crystal frame of the CDP, as shown in Figure \ref{fig:sig3pf}, they are most interpretable with respect to prior literature. The six pole mechanism has six Shockley partial vectors along $\hkl<112>$ Shockley partial directions known to contribute to CTB motion. The four pole mechanism has two  $\hkl<112>$ type and two $\hkl<110>$  type directions, consistent with slip vector analysis in \cite{priedeman2017role}. The three pole mechanisms has two vectors pointing along $\hkl<112>$ directions and one along the $\hkl<110>$ direction (with magnitude smaller than a full dislocation), consistent with a picture of boundary migration mediated by triplets of partial dislocations \cite{Humberson2019}. Prior slip vector analysis for the $(110)$ ITB in \cite{priedeman2017role} originally reported a mechanism with nine displacement types, but this analysis was later amended to a three shuffle mechanism in \cite{bair2019antithermal} upon more careful analysis of thermal noise. 

Exhaustive enumeration of shuffling patterns for $\Sigma3$ GBs via the forward model in \cite{OT} (without shear coupling and at small regularization) showed four distinct classes of transformations with  3, 4, 6 and 7 unique displacement poles for different microscopic translations. The first three types of shuffling patterns are found in our MD data and the relative frequency of these transformations in inclination space matches the frequency in translation space. For instance, a randomly chosen plane inclination or microscopic shift is most likely to yield a 3 shuffle distance minimizing transformation. If the microscopic shift $\bm{p}$ is chosen explicitly in the dichromatic pattern on the basis of energy minimization (as in Fig \ref{fig:sig3pf}), the displacement patterns enumerated by the forward model match the MD data exactly. 

The correspondence between the distance minimizing shuffling pattern selected for different microscopic shifts and the shuffling pattern selected for different plane inclinations has important implications for modeling grain boundary mobility. The problem of determining the dependence of mobility on plane inclination can be approximated as the problem of determining the energetics of a discrete set of shuffling patterns which vary with microscopic shift. Different values of  $\bm{p}$ which are predicted to correspond to distinct transformations can be used to seed bicrystal simulations or NEB calculations. For each characteristic displacement texture, there is a distinguished choice of $\bm{p}$ which minimizes total shuffling distance (and defines the OTDP). It is hypothesized that this choice of shift vector locally maximizes mobility within a characteristic displacement texture class. 

Characteristic displacement textures are observed for all other misorientations with sufficient inclination sampling in the FCC Ni dataset (Figure \ref{fig:sigtrends}). $\Sigma 5-11$ GBs provide a rich and complex set of migration mechanisms for further analysis, but are only pursued in this work to the extent that they illustrate new concepts. Shear coupling behavior and thermal behavior of selected $\Sigma 5$ GBs is discussed in Section \ref{sec:sig5}. The multimodal migration of a $\Sigma 11$ GB is discussed in Section \ref{sec:sig11}, where it is shown that choice of $\bm{p}$ also impacts the relative amount of GB diffusion and shuffling expected during migration. 

\subsubsection{Degenerate displacements and implications for period doubling and sliding}\label{sec:degen}

\begin{figure}[th]\
    \centering\leavevmode
    \includegraphics[width=0.5\textwidth]{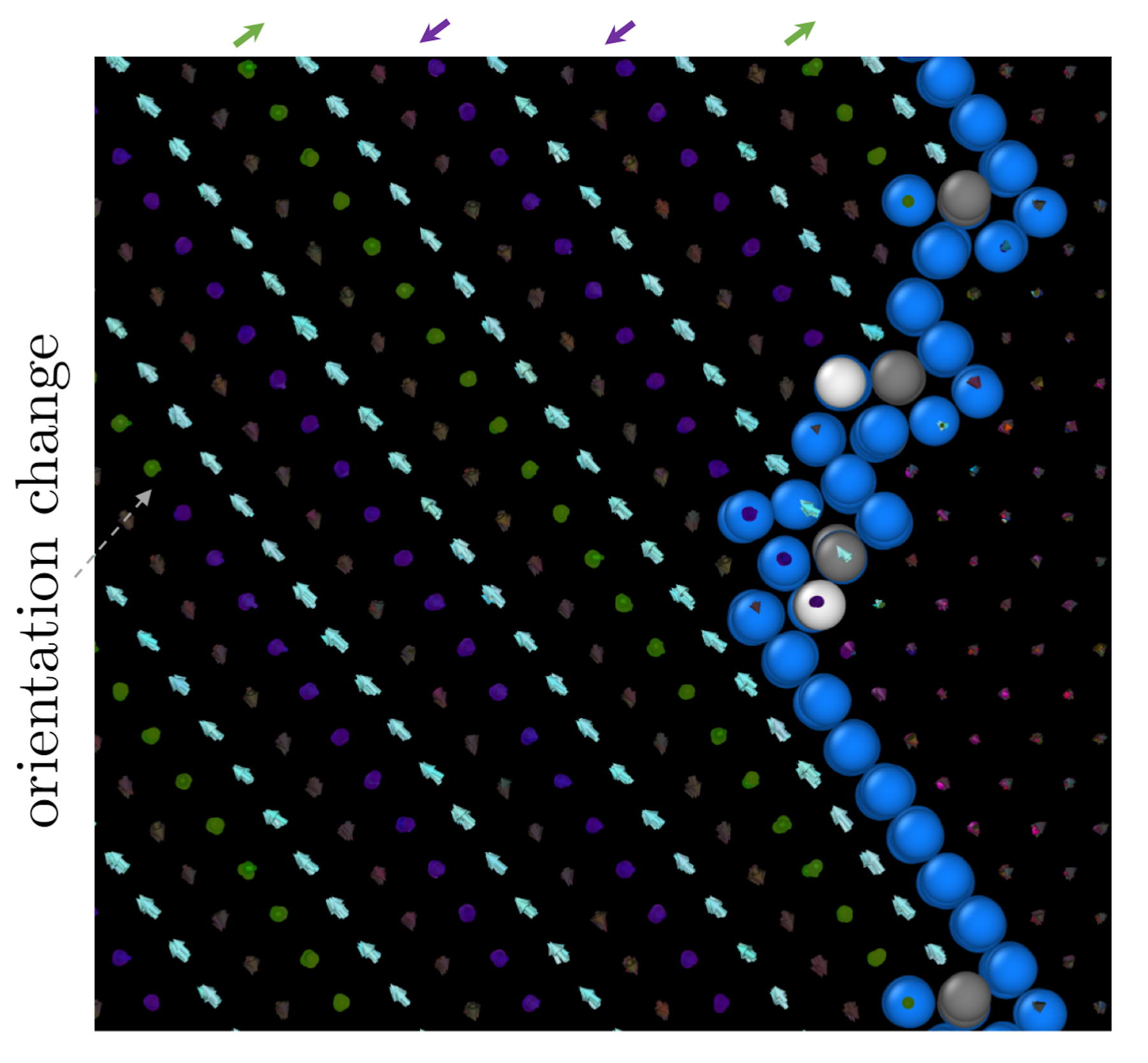}
    \caption{Displacement texture for $\Sigma 3$ $\hkl(811)||\hkl(554)$ asymmetric tilt GB at 300 K demonstrates the 4 pole transformation mediated by \hkl{112} ITB facet migration. The length degeneracy in the displacement pattern results in a period doubling effect where purple and green displacements alternate orientation along layers of \hkl(111) planes. An orientation change within a single plane is highlighted. Such period doubling effects are common across the FCC Ni dataset and are sensitive to microscopic shift $\bm{p}$.}
    \label{fig:sig3super}
\end{figure}

Displacement texture aids the interpretation of the thermal and mechanical behavior of $\Sigma 3$ GBs. An important aspect of the four and six pole transformations for the $\hkl(112)$ ITB and $\hkl(111)$ CTB is that they contain degenerate length displacements in the dichromatic pattern. According to the geometry based forward model, an atom is equally likely to choose any of the degenerate displacements during migration. However, in reality, during migration, once the degeneracy in the shuffling pattern has been broken (by a fluctuation or small external perturbation), the remainder of the atomic displacement must proceed in a way that maintains compatibility of the net reorientation. In the MD data, an interesting period doubling effect is observed for $\hkl(112)$ ITB motion in which every tenth plane of shuffles alternates between degenerate displacements of opposite parity in the 4 pole transformation (purple and green vectors in Figure \ref{fig:sig3super}). It is expected that one displacement type could be favored with strains which break the degeneracy in the dichromatic pattern. 

The degeneracy of the six pole transformation for the CTB has implications for stress generation during migration. Three degenerate twinning partials are available on two of three parallel $\hkl(111)$ planes leading to six unique displacement directions total for sliding (Figure \ref{fig:sig3pf}).  As described by Thomas in \cite{Thomas2019}, all disconnection modes ($\bm{b}$,$\bm{h}$) for the CTB can be obtained from integer multiples of two burgers vectors $\bm{b_1} = \hkl[1-12]a/6$ and $\bm{b_2} = \hkl[-2 1 1]a/6$ (where $a$ is the lattice parameter) and integer multiples of step heights $\bm{h_{10}} = -\hkl[-111]a/3$. The simplest type of disconnection motion involves only one displacement given by one of three twinning partials ($\bm{b_1}$, $\bm{b_2}$ or $\bm{b_3} = -\frac{1}{2}(\bm{b_1} + \bm{b_2}))$ and has step height $\bm{h_{10}}$ associated with the spacing between neighboring $\hkl(111)$ planes. This type of motion is favored by a strain along $\bm{b_1}$, $\bm{b_2}$ or $\bm{b_3}$. It is an example of a pure shear transformation with no shuffles. Energy jump driven migration of the CTB in our simulations is associated with only a small amount of net shear, even for significant normal motion. The low coupling factor of this transformation, combined with the presence of all six displacements, indicates an alternating mixture of sliding displacements during migration. Thus, even though the GB may move with little apparent shear, it generates a shear stress and strain in its transition state. This behavior is contrary to the prediction of the disconnection model that energy jump driven motion of the CTB corresponds to pure step motion. Pure step motion is not possible because of the sliding degeneracy in the min-shuffle mapping. Stresses are generated during constrained migration of the CTB despite the zero net shear character of motion observed in many of our simulations. Such a situation involving alternating sliding events is likely common during zero shear migration of ordered GBs \cite{hirth2016disconnections} and has been reported in atomistic simulations which show anomalous stress driven motion of  $\Sigma 7$ GBs with zero net strain \cite{molodov2011migration,wan2017atomistic}. We note that alternating sliding events which result in migration with zero apparent shear may be coupled to shuffling (with finite coupling factor) or may have pure sliding character (infinite coupling factor).

The $\Sigma 3 \hkl(112)$ ITB generates a stress under constraints for different reasons: zero shear migration is in competition with partial dislocation emission, a process associated with a small shear strain \cite{hirth2016disconnections}. At sufficiently low temperatures with free boundary conditions ($T < 900$ K) and a ramped energy jump driving force, the $\hkl(112)$ ITB emits stacking faults into the grain ahead of the interface, resulting in a net shear. In the constrained case, this shear generates a back stress which arrests further dislocation extension. Migration is then promoted as driving force is ramped upward. Displacement poles may deviate from their ideal positions in the characteristic displacement texture with changing length of emitted dislocation and with changes in the partial dislocation structure of the GB. Extended defects may point outward or inward from the GB plane depending on temperature. The OTDP is a convenient representation in this case because small strains that arise from changes in boundary structure and dislocation emission are subtracted out of the displacement texture. The displacement pole locations in the OTDP vary little with temperature from 100 to 1000 K, though the relative probabilities of degenerate displacements may vary. 

\subsubsection{Quantifying thermal noise for high mobility migration of the $\Sigma 3 \hkl(110)$ GB}\label{sec:sig3110}

 \begin{figure}[th]\label{sig3thermal}
    \centering\leavevmode
    \includegraphics[width=0.7\textwidth]{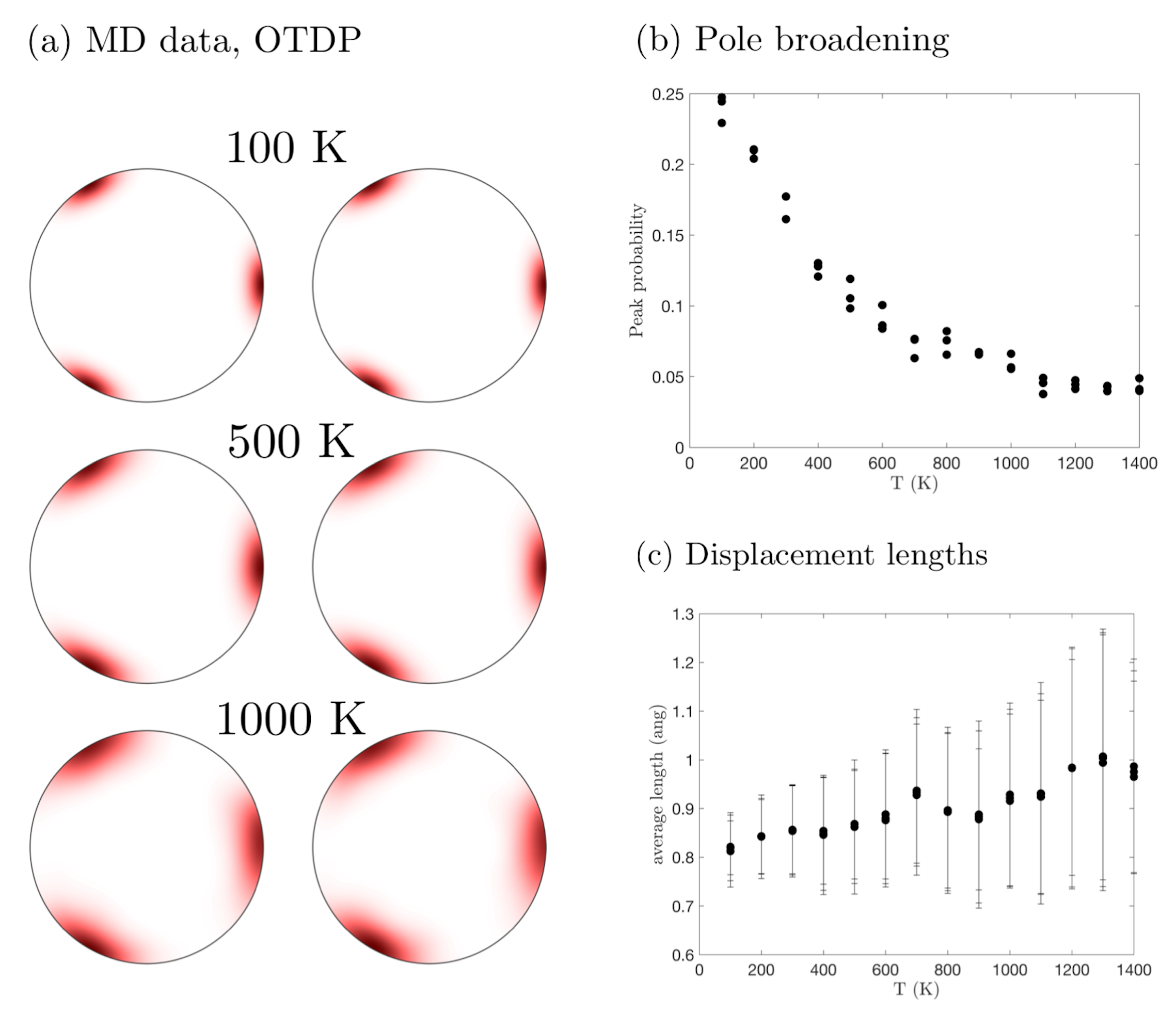}
    \caption{Displacement texture analysis allows for quantification of thermal behavior. (a/b) Displacement poles for $\Sigma 3$ \hkl(110) GB broaden with temperature, as shown by a decrease in maximum pole probabilities. (c) Displacement lengths also tend to increase with temperature. The analysis is performed in the OTDP to regularize over small shifts in displacement fields arising from the Shockley partial structure of the GB.}
    \label{fig:sig3thermal}
\end{figure}

To illustrate the detailed information that can be extracted from displacement textures, we characterize the thermal behavior of the highly mobile $\Sigma 3$ \hkl(110) ITB at temperatures from 100-1400 K. The three pole transformation remains stable at high temperatures and the probability mass around each pole broadens (Figure \ref{fig:sig3thermal}a). One way to quantify thermal broadening of the displacement texture is by tracking the maximum probability of each pole in the spherical harmonic fit. Displacement length distributions can also be computed for the displacements corresponding to each pole. It is observed that the angular deviation of shuffles and the average length of shuffles generally increase with temperature, matching intuition. However, the average increase in shuffle length and the decrease in peak probability are not necessarily monotonic. Small changes in the Shockley partial structure of the boundary may impact thermal broadening of the displacement texture. The probability decay is not well modeled by a gaussian solution to the diffusion equation. 

\subsection{Characterizing the misorientation dependence of displacement texture}\label{sec:geotrends}

In this section, we characterize displacement textures for mobile GBs at a variety of misorientations including both low and high $\Sigma$ GBs. At 300 K, over a third (142/388) of the GBs in the FCC Ni dataset move with near zero shear ($|\beta| < 0.05$) under unconstrained boundary conditions and a ramped energy jump driving force \cite{yusurvey}. We begin by characterizing the motion of low $\Sigma$ GBs which move with zero net shear. The correspondence between plane inclination and microscopic shift for $\Sigma 3$ GBs is found to generalize to $\Sigma 5-11$ GBs. Next, we show that shuffling patterns may retain a high degree of order even for high $\Sigma$ GBs. We discuss common features of displacement textures for high $\Sigma$ GBs such as vortex displacement patterns. 

\subsubsection{Characteristic shuffling patterns for low $\Sigma$ GBs}\label{sec:bpaniso}

\begin{figure}[th]
    \centering\leavevmode
    \includegraphics[width=0.3\textwidth]{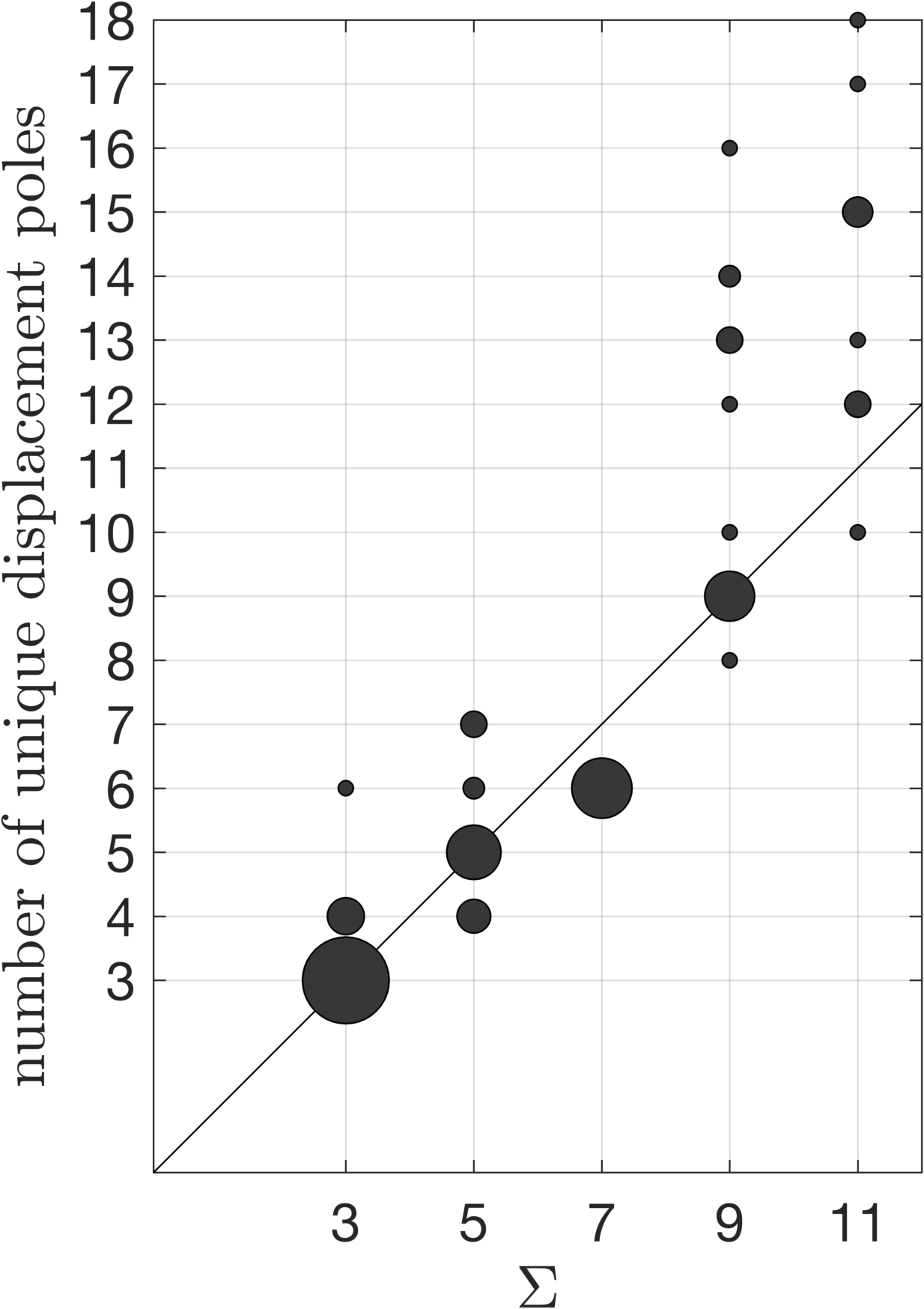}
    \caption{Distinct characteristic displacement textures at fixed misorientation can be separated into classes by the number of unique shuffle displacement types in the displacement pole figure. Point size is proportional to the number of boundaries within each class.}
    \label{fig:sigtrends}
\end{figure}

Different characteristic displacement textures are observed for distinct misorientations in the FCC Ni dataset depending on boundary plane inclination. The number of unique displacement poles (tabulated for constrained migration of low $\Sigma$ GBs at 300 K and shown in Figure \ref{fig:sigtrends}) is found to be a useful identifier of distinct characteristic displacement textures. For each misorientation, a \textit{minimal} characteristic displacement texture is observed with the fewest shuffle displacement types. The minimum number of shuffles is found to equal $\Sigma$ for $\Sigma 3$ GBs and $\Sigma-1$ for $\Sigma 5-11$ GBs. These values can be rationalized by the CSL geometry of the CDP. If one atom in every CSL unit cell is coincident, $\Sigma-1$ displacement types are required for reorientation. However, depending on crystal structure, the CDP may have degenerate length shuffles which make observation of more than $\Sigma-1$ shuffles likely. This is the case for $\Sigma 3$ GBs, where the zero shear transformation in the CDP is associated with the six shuffle mechanisms discussed for the \hkl(111) CTB. In $\Sigma 3$ GBs in BCC Fe, such degeneracy is not present in the CDP and 2-shuffle transformations are common. Depending on microscopic shift, different types of degeneracies in shuffling patterns may arise such that (min(\# poles ) $>$ $\Sigma-1$). These boundary plane dependent shuffling transformations are observed for all low $\Sigma$ GBs in Figure \ref{fig:sigtrends} except $\Sigma 7$ GBs. We note that some GBs that have a large number of shuffles may also contain non-optimal displacements (as discussed in Sections \ref{sec:sig5}, \ref{sec:hisig} and \ref{sec:multi}).

Shear coupling changes the expected shuffling geometry in the dichromatic pattern and can reduce the number of shuffles required for reorientation to well below $\Sigma - 1$ and even zero in the case of a pure shear transformation. For example, the shuffling pattern enumerated in the forward model for a $\Sigma 15 \hkl<112>$ tilt GB has only four unique shuffle vectors  (Figure \ref{fig:helix}), fewer than the 14 expected on the basis of analysis of zero shear transformations. 

 \begin{figure}[th]
    \centering\leavevmode
    \includegraphics[width=0.8\textwidth]{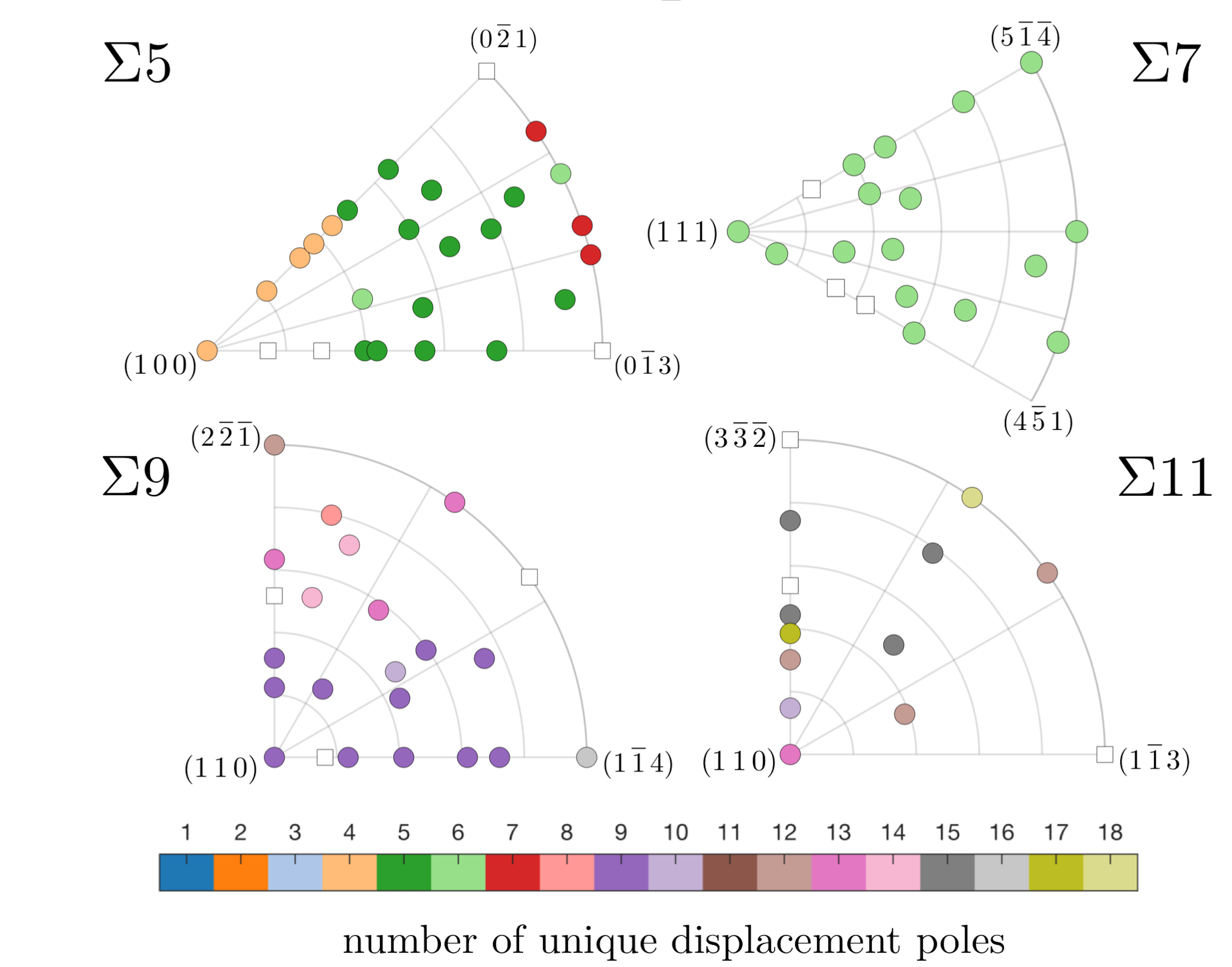}
    \caption{Distinct zero shear characteristic displacement textures at fixed misorientation can be separated into classes by the number of unique shuffle displacement types in the displacement pole figure. Plotted squares denote boundaries which sweep out shear during motion and are excluded from the current analysis.}
    \label{fig:bpaniso}
\end{figure}

In light of the analysis of $\Sigma 3$ GBs in Section \ref{sec:sig3}, we hypothesize that the distinct classes of shuffling transformations that arise for different plane inclinations in $\Sigma 5-11$ GBs can be explained by the impact of microscopic shifts on distance minimizing mappings in the dichromatic pattern. The forward model may be used to test this hypothesis by exhaustively enumerating characteristic displacement textures with respect to shift vector and regularization and comparing to MD data. In the next section, we compare $\Sigma 5$ displacement textures observed in MD data to predictions of the forward model. In Section \ref{sec:sig11}, this analysis is extended to better understanding the rigid sliding and migration of a $\Sigma 11$ twist GB. 

\subsubsection{Characteristic shuffling patterns for $\Sigma$5 GBs}\label{sec:sig5} 

$\Sigma 5$ GBs exhibit diverse shuffling behavior at 300 K with at least four distinct transformation types observed under constrained boundary conditions with varying plane inclination (Figure \ref{fig:bpaniso}). The four pole transformation corresponds to the $\Sigma 5 \hkl(100)$ twist GB and  $\Sigma 5$ GBs with nearby inclinations. This four-shuffle mechanism was observed in early MD studies of curvature driven GB migration by Jahn and Bristowe \cite{bristowe} and is optimal in the sense that it minimizes net shuffling distance in the TDP \cite{OT}. Five, six and seven pole transformations are also observed for distinct inclinations (Figure \ref{fig:bpaniso}). Several GBs (marked with squares in Figure \ref{fig:bpaniso}) slide significantly even under constrained boundary conditions. 

 \begin{figure}[th]
    \centering\leavevmode
    \includegraphics[width=0.7\textwidth]{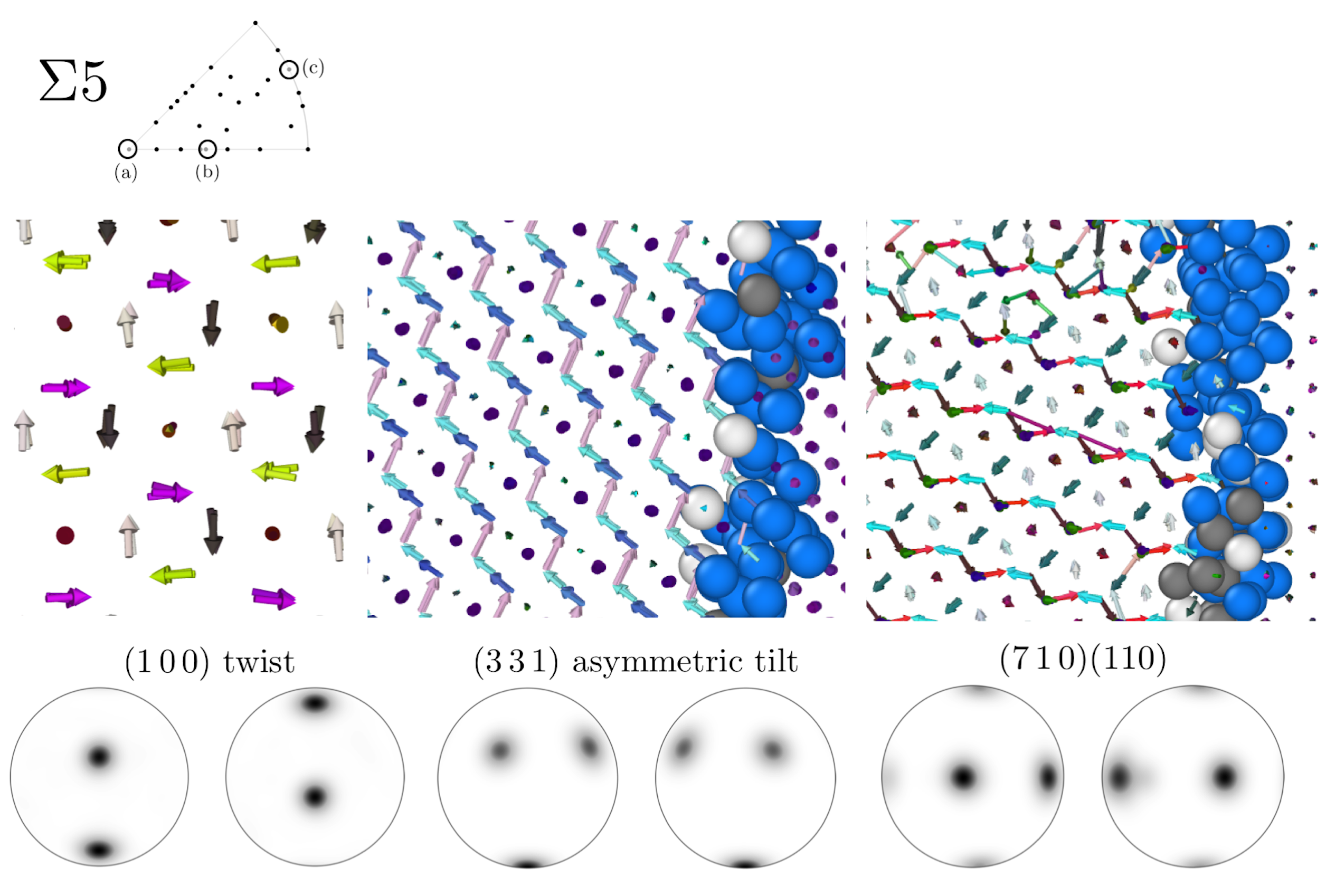}
    \caption{Three types of shuffling patterns observed for $\Sigma 5$ GBs with zero net shear deformation and containing (a) 4, (b) 5 and (c) 7 or more nonzero poles. All snapshots are shown at 500 K.}
    \label{fig:sig5zs}
\end{figure}

 \begin{figure}[th]
    \centering\leavevmode
    \includegraphics[width=0.7\textwidth]{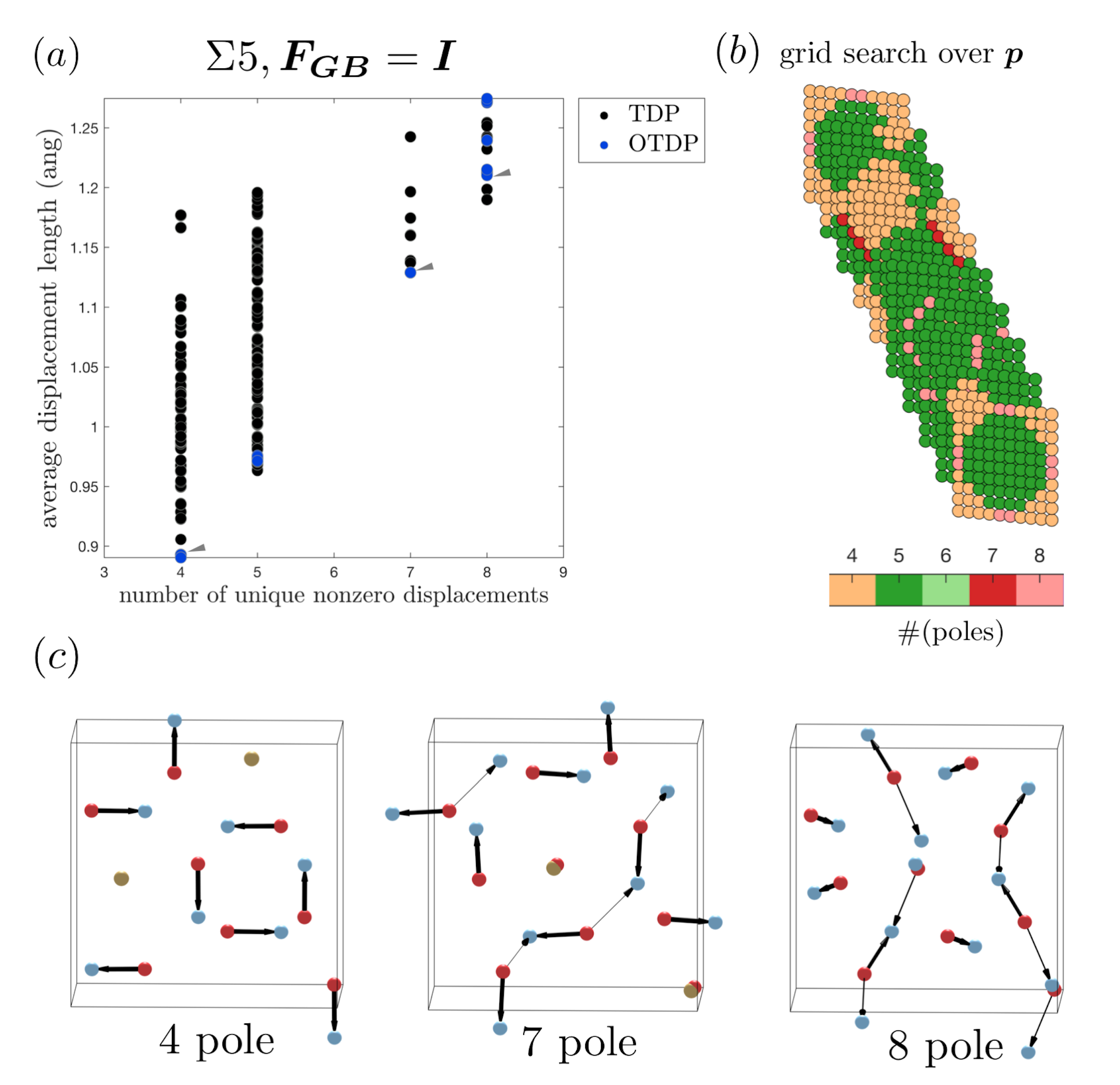}
    \caption{Enumeration of shuffling patterns for $\Sigma 5$ GBs in forward model with $\bm{F^{GB}} = \bm{I}$, $\epsilon = 0.005$ and varying microscopic shift $\bm{p}$. (a) Grid search over shift vector in DSC lattice reveals several distinct transformations, conveniently parametrized by number of unique nonzero displacements. (b) Transformations depend on the choice of $p$ within the search space in distinct ways. (c) Three example transformations are shown. The last transformation is particularly significant because of its appearance in both the roughened migration of the $\Sigma 5$ \hkl(310) GB \cite{OT} and zero shear migration of asymmetric tilt GBs.}
    \label{fig:sig5ot}
\end{figure}

Non-minimal zero shear characteristic displacement textures for $\Sigma 5$ GBs may be rationalized by the impact of the microscopic shift vector on allowed mappings in the TDP, similar to prior analysis for $\Sigma 3$ GBs. The forward model is used to enumerate shuffling patterns for 1000 shift vectors in the $\Sigma 5$ DSC unit cell with $\bm{F^{GB}} = \bm{I}$. At least four distinct transformations are observed across both the MD data and the forward model. Transformations from the forward model and MD data show striking similarities. For instance, the 8 pole transformation in Figure \ref{fig:sig5ot}, which arises from degenerate displacements in the TDP, appears to match the displacement texture in Figure \ref{fig:sig5zs} for a $\Sigma 5$ asymmetric tilt GB up to a rigid rotation. Interestingly, a very similar pattern was observed during roughened migration of the $\Sigma 5$ \hkl(310) GB at high temperatures in \cite{OT}, where it is argued that the pattern arises from a combination of disconnection modes of opposite sign along with longer, non-optimal vectors. The forward model offers a way to simplify and explore the rich phase space of GB motion. The two highest displacement numbers in the forward model $(7,8)$ and the MD data $(6,7)$ do not match. This mismatch arises from how poles in the MD data are counted in the presence of noise. The unique pole counts tabulated from the MD data should be viewed as a lower bound for the true number of unique displacement types because 1) poles must be distinguished within an angular cutoff in the MD data and 2) low probability displacement types below a threshold probability are ignored. With further work, such as the templating method described in \cite{OT}, displacement textures from MD data and the forward model can be compared more quantitatively. 

\subsubsection{Characteristic shuffling patterns for high $\Sigma$ GBs}\label{sec:hisig}

In this work, a GB is called high $\Sigma$ if local maxima in the displacement texture are sufficiently close in angle to form continuous bands in the displacement pole figure. 
This typically occurs for $\Sigma > 19$ within our dataset. Previously, it has been assumed that high $\Sigma$ GBs cannot move via the ordered shuffling transformations of low $\Sigma$ GBs because of the complex shuffles required in large CSL unit cells (\cite{Yan2010}).  The FCC Ni migration dataset offers several counterexamples to this statement, but also highlights the importance of simultaneous GB diffusion and migration mechanisms during the motion of high $\Sigma$ GBs. 

\begin{figure}[th]
    \centering\leavevmode
    \includegraphics[width=0.7\textwidth]{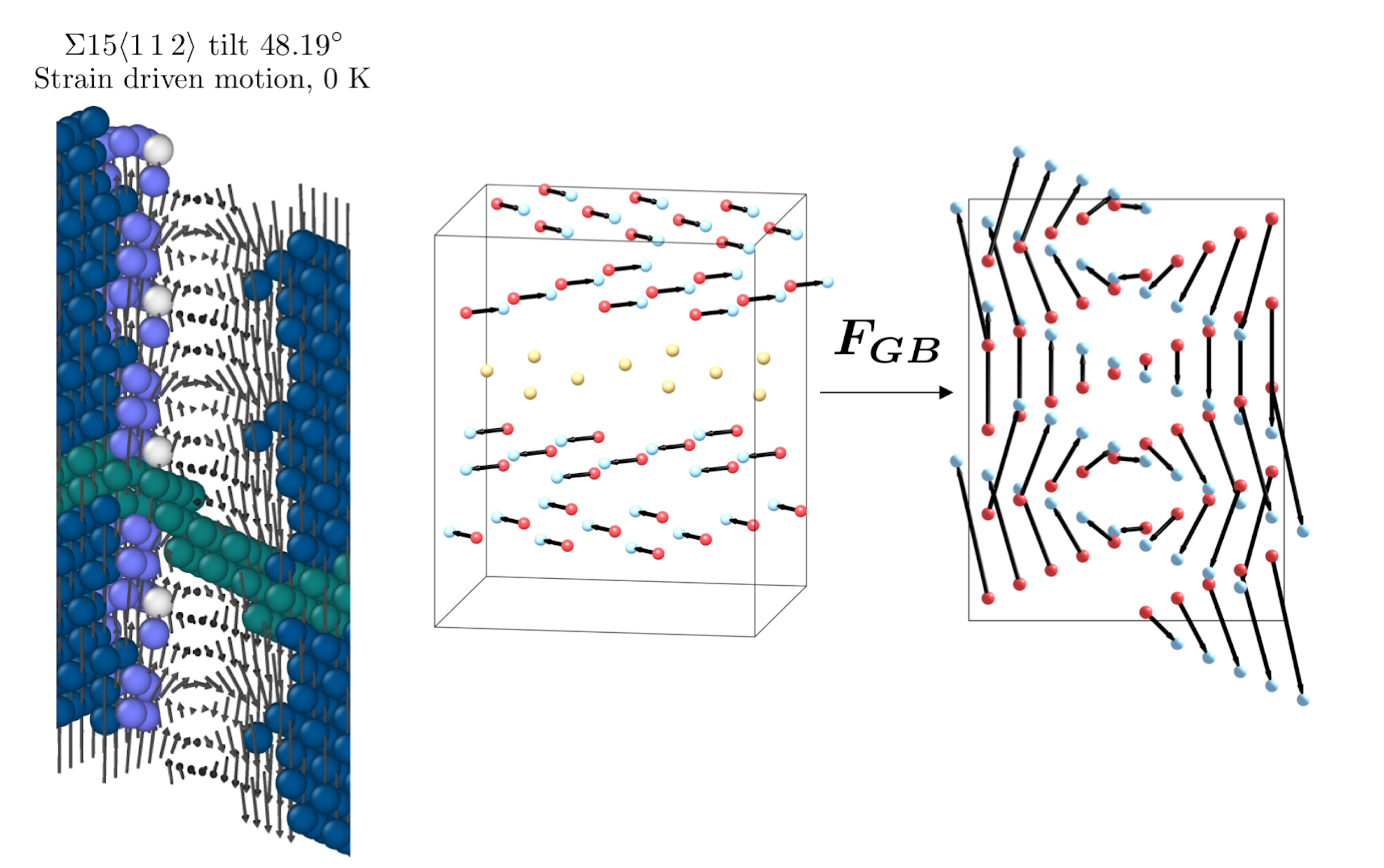}
    \caption{The displacement texture of the $\Sigma 15$ \hkl<211> $48.19^\circ$ tilt GB can be decomposed into a set of layered shuffles (middle) and a shear deformation associated with a zero shear reference angle of $0^\circ$. The helical displacement texture (left and right panels for predictions of MD and the forward model) results from the superposition of shuffles and shears.}
    \label{fig:helix}
\end{figure}

Two ubiquitous ordered displacement morphologies during high $\Sigma$ GB migration include \textit{displacement vortices} and \textit{displacement helices}. These patterns often arise during shear coupled migration of symmetric tilt GBs (Figure \ref{fig:helix}). Vortices and helices accommodate a net shear deformation and can be decomposed into ordered layers of shuffles and pure shear displacements. Vortices may also be present in systems with zero net shear deformation, such as twist GBs. In twist GBs, displacement vortices are confined to the interface plane, circulating around an axis $\hat{\bm{v}}$ pointing in the GB normal direction. In symmetric tilt GBs, vortices circulate around the tilt axis $\hat{\bm{v}}$ in planes perpendicular to the GB plane. In asymmetric tilt and mixed tilt/twist GBs, vortices may lie along planes inclined to the GB. Interestingly, the length scale of vortex displacement patterns is sensitive to geometric confinement. \textit{Basketweave} displacement textures are observed during the constrained migration of symmetric tilt GBs in which vortices tile space with zero net shear. These patterns, discussed further in Section \ref{sec:st}, are hypothesized to arise from a primary disconnection mode mixed with one or more mechanisms that reduce the net shear deformation, such as as GB diffusion coupled sliding or secondary disconnection motion. 

\begin{figure}[th]
    \centering\leavevmode
    \includegraphics[width=0.5\textwidth]{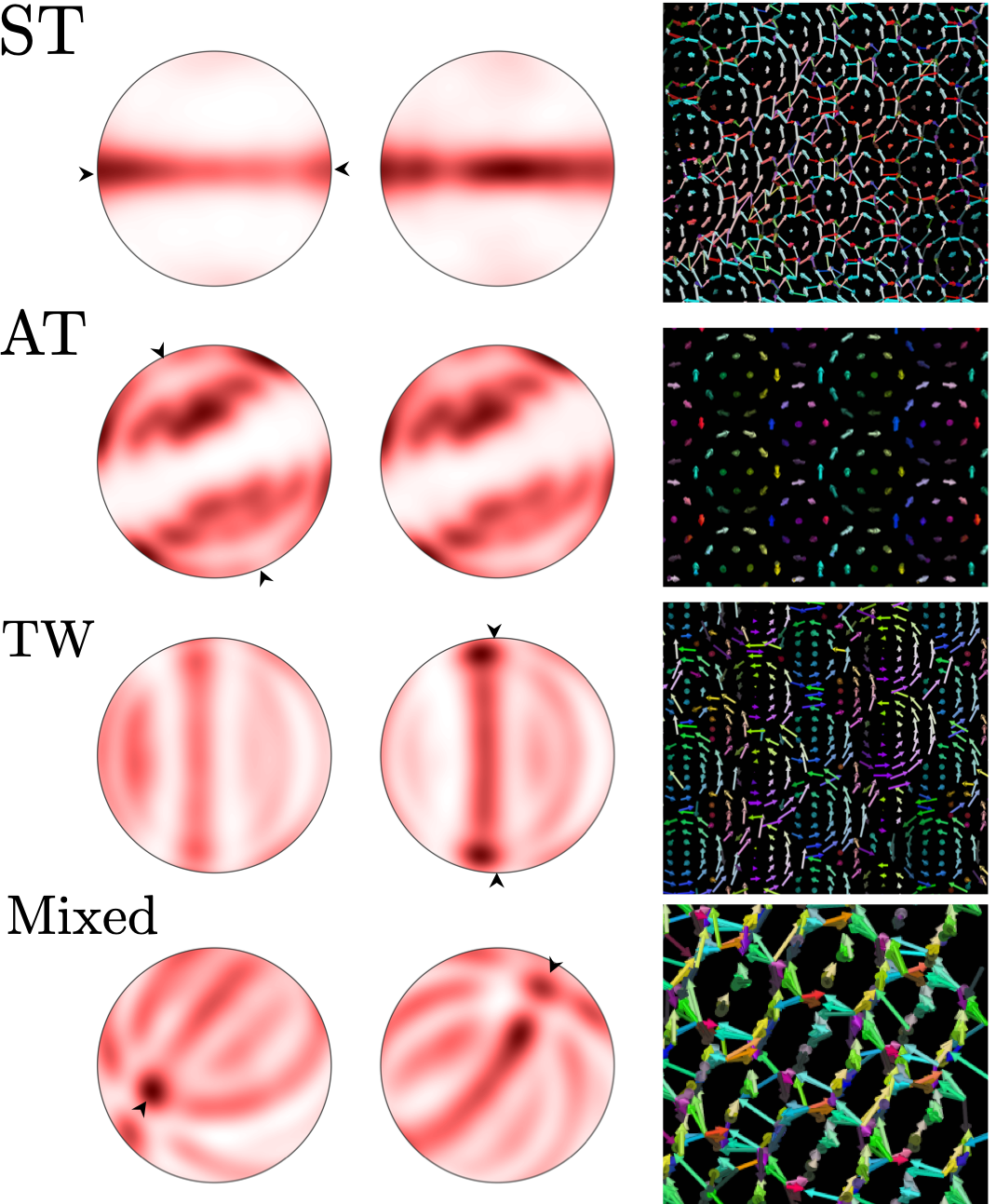}
    \caption{Displacement pole figures for high $\Sigma$ GBs of varying GB type (ST = symmetric tilt, AT = asymmetric tilt, TW = twist, Mixed = mixed tilt/twist). Arrows on displacement pole figures show the planar cut that is used to visualize the shuffling patterns. Cross sections are perpendicular to the tilt axis and in the plane of the boundary for ST and TW GBs. GBs are ST: $\Sigma 89$ \hkl<100> 28$^\circ$ (ID 264), TW: $\Sigma 235 \hkl(310)$ (ID 369), AT: $\Sigma 51 \hkl(521)$ (ID 362), Mixed: $\Sigma 81 \hkl(8 4 1) || \hkl(7 4 4)$ (ID 273), where IDs are the same as in \cite{olmsted2009survey}.} 
    \label{fig:sighi}
\end{figure}

It is demonstrated via the forward model that the ordered layers of shuffles extracted from the shear coupled migration trajectories in Figure \ref{fig:helix} are optimal in the sense that the shuffles minimize net transformation distance in the TDP. In the FCC Ni dataset, unimodal shear coupled migration at low and intermediate temperatures (100-1000 K) satisfies the min-shuffle hypothesis. Nevertheless, a small fraction of non-optimal displacements may be present that grow in probability with increasing driving force or temperature. Non-optimal displacements correspond to competing atomic rearrangement mechanisms. For instance, the $\Sigma 89$ \hkl<100> $26^\circ$ tilt GB exhibits diffusional flow along the tilt axis during migration that increases in probability with temperature. The displacements along the tilt axis appear in the displacement pole figure as poles isolated from the continuous band of vortex displacements (Figure \ref{fig:sighi}). Depending on GB geometry (both microscopic and macroscopic degrees of freedom $\bm{p}$), non-optimal displacements may form networks. Large, non-optimal displacements of the $\Sigma 85$ \hkl(100) twist GB were previously found to form rectangular arrays within the plane of the GB \cite{OT} and are thought to be associated with GB self diffusion in screw dislocation cores. Non-optimal displacements were localized to regions where minimum length shuffles were large. The $\Sigma 85$ \hkl(100) twist GB is an example of a GB in the FCC Ni dataset that always has a relatively high fraction of non-optimal displacements, even at low and intermediate temperatures. Because of the long shuffles necessary for geometric compatibility, and the likelihood of many pathways of similar length, many high $\Sigma$ GBs are dissimilar from special low $\Sigma$ GBs which can move via athermal shuffling mechanisms that do not involve diffusion. Shear coupling reduces the complexity of shuffles but may still involve significant amounts of GB diffusion, depending on microscopic shift, temperature and driving force. 

\subsection{Characterizing nonuniform and roughened migration}\label{sec:multi}

During grain growth, boundaries do not migrate uniformly. Migration is often observed to be a jerky, stochastic process associated with changes in local driving forces and boundary conditions. It is important to gain a better fundamental understanding of nonuniform GB migration. A counterpart to nonuniform migration is roughened migration. Above a critical roughening transition temperature that depends on GB type and active migration mechanism, GB mobility jumps from near zero to a larger value. A recent disconnection based model for roughening points to the dual roles of temperature and driving force in facilitating roughening (\cite{chen2020grain}). High driving forces (energy jumps or stresses) lower the effective roughening temperature. During ramped driving force MD simulations of bicrystals, transitions in migration behavior are observed which depend on GB type, temperature, driving force and boundary conditions \cite{yusurvey}. In this section, we employ ramped driving force simulations in bicrystals with constrained boundary conditions. With this simulation setup, GBs which yield with sliding character undergo up to two regimes of motion: 1) an initial slow phase of motion characterized by stair-steps in the GB position and average shear stress and 2) a fast period of continuous runaway motion that accommodates zero shear boundary conditions. We denote the driving force at the initial onset of motion as the \textit{critical driving force} and the driving force at the onset of runaway motion the \textit{second critical driving force}. A mechanistic account of roughening at the second critical driving force is a main contribution of this section. 

Two types of nonuniform migration will be examined in this section. First, we will explore the sequential sliding and migration of a $\Sigma 11$ twist GB. This GB is representative of a poorly understood behavior for twist GBs observed in a recent MD survey \cite{yusurvey} in which GBs rigidly slide by a small increment before moving forward. $\Sigma 11$ GB motion has also recently been analyzed in the context of asymmetric motion in which a specific GB was found to have different mobilities in the forward and backward direction under energy jump driving forces of opposite sign \cite{mccarthy2020shuffling,mccarthy2021alloying}. In Section \ref{sec:st}, we explore pattern formation during roughened migration of symmetric tilt GBs upon second yield. An open question in the literature concerns the mechanisms by which GBs lose their shear coupling factor at high temperatures or under constraints. Zig-zag motion has been observed in a specific case \cite{thomas2017reconciling}, but it is not well understood whether zig-zag motion is generic to roughened migration or whether other mixed mechanisms such as diffusion assisted sliding and coupling are possible. We show that characteristic tiled vortex patterns during constrained migration of tilt boundaries arise from a combination of disconnection motion, sliding and GB diffusion. Depending on GB geometry, zig-zag motion may also be observed. These examples demonstrate how displacement texture analysis, combined with a forward model for migration, assist the interpretation of multimodal migration data.  

\subsubsection{Nonuniform sliding and migration of a $\Sigma 11$ twist GB}\label{sec:sig11}

\begin{figure}[th]
    \centering\leavevmode
    \includegraphics[width=0.7\textwidth]{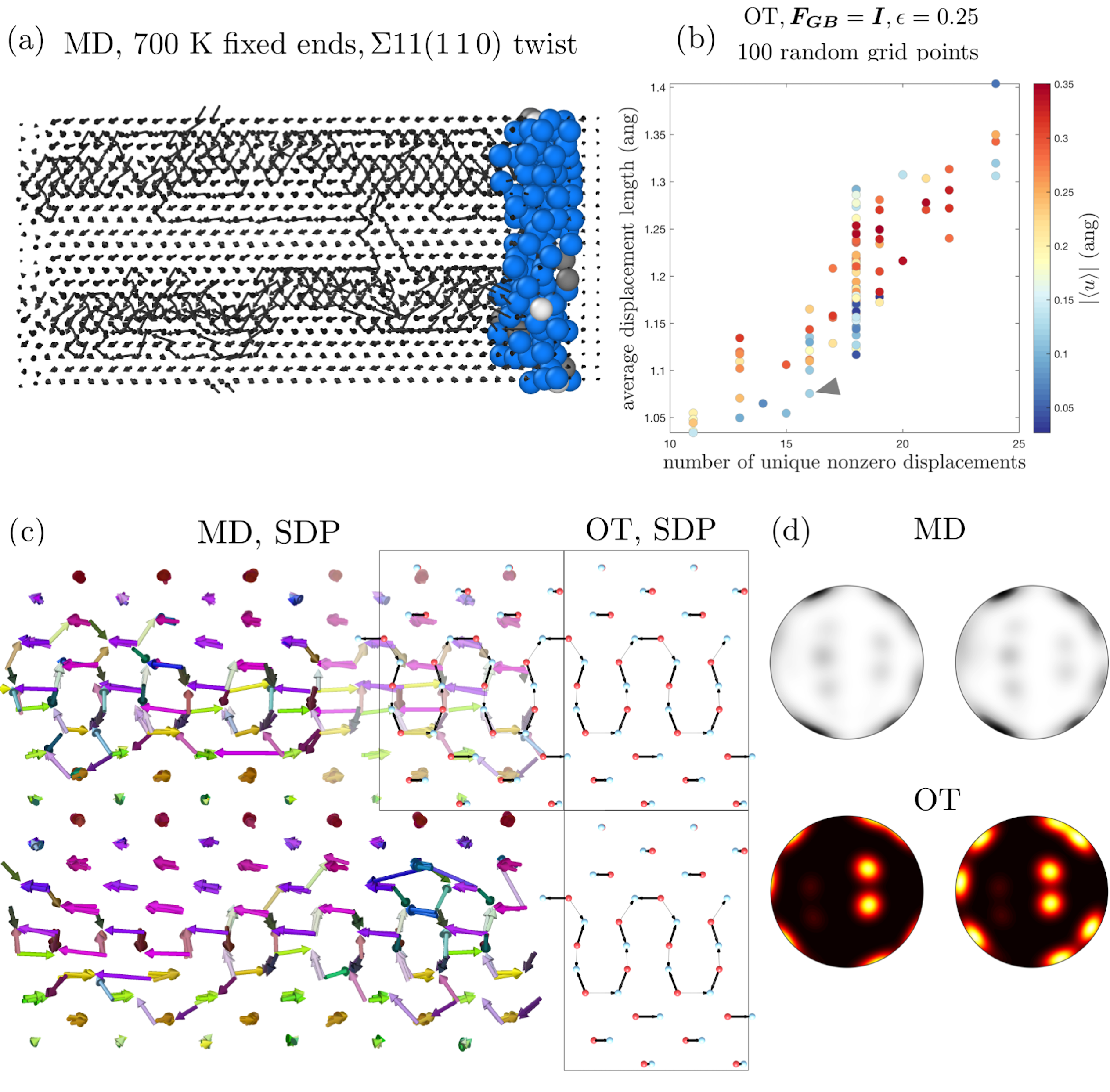}
    \caption{(a) Displacement texture in TDP for $\Sigma 11$ \hkl(110) twist GB demonstrates banded displacement pattern. Because a sliding event occurs at the onset of motion, the TDP is modified from the initial state and is treated as unknown. (b) Shuffling patterns are enumerated for 100 random shifts $\bm{p}$ at relatively large regularization $\epsilon = 0.25$. (c/d) The best match to MD data is found to be a 16 pole transformation. Though the layered structure is apparent in the prediction of the forward model, the MD data clearly contains more types of displacements. This example shows the general strategy of optimizing over $\bm{p}$ and $\epsilon$ in order to match the MD data.} 
    \label{fig:sig11}
\end{figure}

The $\Sigma 11$ \hkl(110) GB offers an example of multimodal migration that involves rigid sliding at the beginning of motion followed by normal migration. In combination with the forward model, displacement texture analysis gives insight into shuffling geometry during this type of motion. At least 17 unique shuffles are present, indicating a non-minimal displacement texture. Bands of displacements are visually apparent with relatively large vertical components along the Y direction (Figure \ref{fig:sig11}). Ordered layers of shuffles lie in between these bands with components primarily in the XZ plane. The displacement pole figure in Figure \ref{fig:sig11}d distinguishes between the displacements in the bands and the displacements between bands along the equator. 

This displacement geometry can be rationalized on the basis of the forward model only if the impact of sliding on microscopic degrees of freedom is taken into account. Additionally, regularization in the forward model is necessary to approximate non-optimal displacements observed in the MD data. Unlike previous examples, the forward model does not reproduce the observed displacement pattern at any $\epsilon$ when $\bm{p}$ is chosen to match the relaxed 0 K bicrystal structure. We hypothesize that a different choice of $\bm{p}$ leads to a matching displacement pattern for some $\epsilon$. A random search over 100 $\bm{p}$ grid points at $\epsilon = 0.25$ reveals a candidate displacement pattern from the forward model as shown in Figure \ref{fig:sig11}b-d. This purely geometric prediction captures the banded pattern shown in the MD data but misses some displacements within the bands compared to the MD data. As the forward model does not capture the details of metastable GB structure and only approximates the net result of multi-hop sequences of displacements, it is unsurprising that it does not predict all non-optimal displacements in the MD data. Nevertheless, it has been shown that alternating shuffling and sliding events accommodate constrained migration of the $\Sigma 11$ \hkl(110) GB. The forward model gives useful insight into the impact of $p$ on allowed mappings in the dichromatic pattern. 

\subsubsection{Constrained migration mechanisms of symmetric tilt GBs}\label{sec:st}

Constrained migration of symmetric tilt GBs often involves basketweave displacement textures with tiled displacement vortices and helices. See, as examples, Figure \ref{fig:intro} (top left) and Figure \ref{fig:sighi} (top). We hypothesize that these patterns can be understood as arising from a primary disconnection mode mixed with a mechanism that reduce the net shear deformation, such as as GB diffusion mediated sliding or secondary disconnection nucleation and motion. Displacement texture analysis aids the interpretation of these complex multimodal migration scenarios.

\begin{figure}[th]
    \centering\leavevmode
    \includegraphics[width=0.5\textwidth]{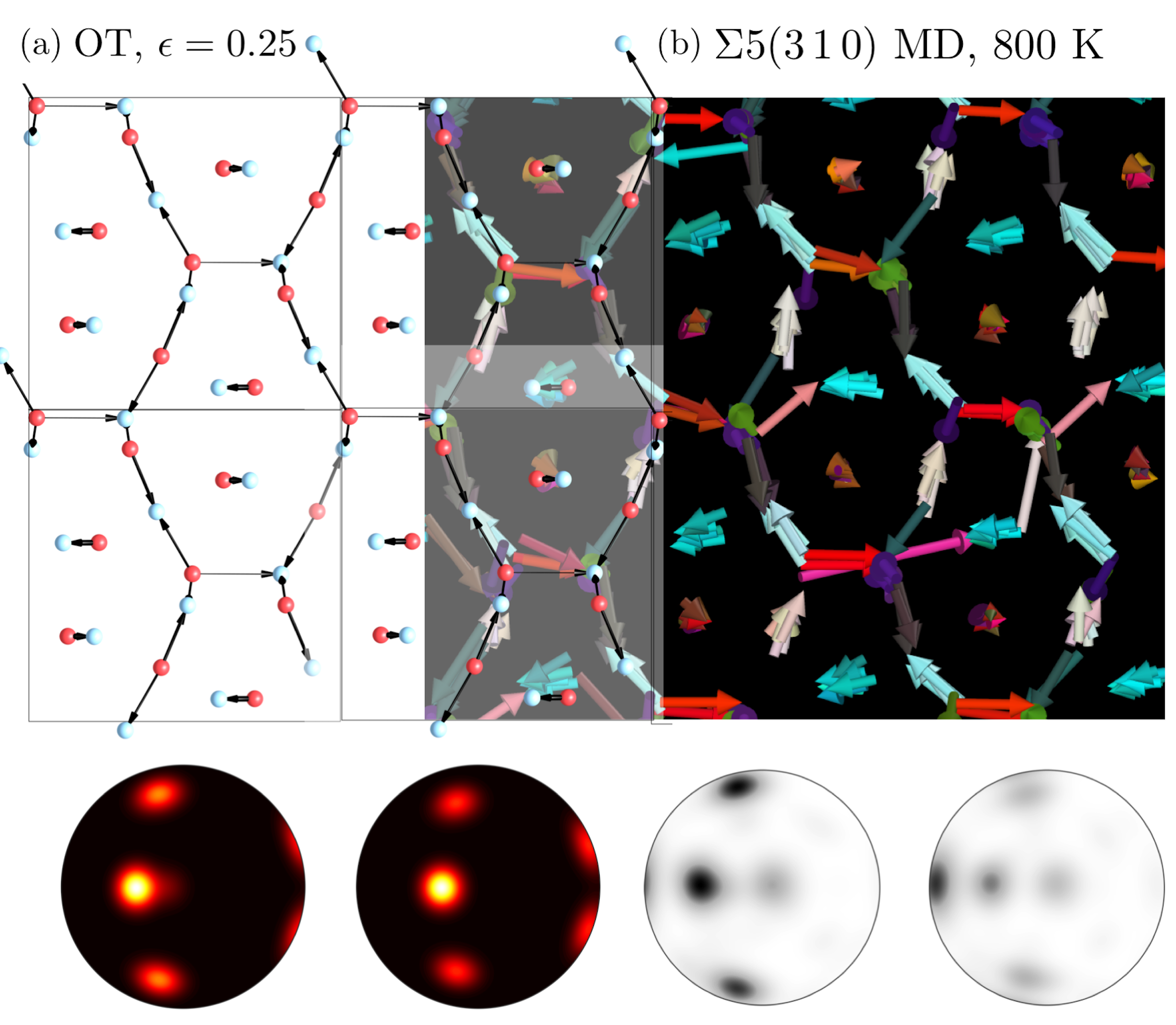}
    \caption{Comparison of displacement textures for (b) roughened migration of $\Sigma 5$ \hkl(310) GB at 800 K under constrained boundary conditions and (a) zero shear transformation enumerated from the forward model with 9 unique poles (regularized version of 8 pole transformation in Figure \ref{fig:sig5zs}).} 
    \label{fig:sig5r}
\end{figure}

In this section, we analyze roughened migration of the $\Sigma 5$ \hkl(310) GB. At temperatures above the roughening temperature, this GB is observed to move with zero net shear regardless of boundary conditions \cite{schratt2020efficient, OT}. Since $\Sigma$ is small, distinct displacements can still be resolved in the displacement pole figure (Figure \ref{fig:sig5r}b). The displacement texture is not minimal and has at least 9 unique nonzero displacements. We attempt to infer the mixture of migration mechanisms from the mixture of displacements in the pole figure. Two candidate mixtures are 1) a laminated combination of two disconnection modes or 2) a zero shear transformation with regularization corresponding to GB diffusion like displacements. 

Displacements corresponding to shear coupled motion of the $\Sigma 5$ \hkl(310) GB in the positive and negative $\hkl(100)$ and $\hkl(110)$ type coupling modes were previously enumerated at small $\epsilon$ \cite{OT}. It was found that the roughened displacement pattern contains displacements from both types of disconnection modes and that shuffles for different modes mix in the plane of the GB, rather than in a zig-zag fashion. However, the mixture predicted from the forward model does not exactly match the MD simulations. Displacements along the tilt axis are observed in the MD data but are not predicted for individual disconnection modes in the forward model \cite{OT}, even at $\epsilon = 0.25$. On the other hand, a zero shear displacement texture predicted from the forward model with appropriately chosen $\bm{p}$ and $\epsilon = 0.25$ captures at least six of the nine poles in the MD data, including displacements along the tilt axis. The agreement is not perfect and the zero shear transformation appears to miss several vortex related displacements in the MD data which \textit{are} captured in the disconnection shuffling predictions. In summary, both of the candidate mixtures of mechanisms from the forward model show partial agreement with the displacement texture from the MD data. We cannot rule out the possibility that there is a different combination of $\bm{p}$ and $\epsilon$ (possible related to microsliding events during migration) in the forward model that give better agreement to the MD data. Alternatively, the simplified distance based potential in the forward model may be insufficient to capture the complexity of the MD simulation, which capture a combination of GB diffusion and disconnection based shuffling with statistics different from the forward model. 

\section{Summary and Conclusions}\label{sec:D}

Shuffling patterns during GB migration strongly depend on bicrystallography, even in complex multimodal migration scenarios. This work provides a simplified modeling and characterization framework for rationalizing the impact of GB geometry in the dichromatic pattern on observed atomic rearrangements during GB motion. High throughput MD simulations in FCC Ni at low and intermediate temperatures reveal a rich variety of GB migration mechanisms which include competing shuffling modes and mixtures of shear coupling, sliding and GB self diffusion. 

The regularized optimal transport based forward model for migration presented in \cite{OT} is shown to significantly clarify analysis of the many mechanisms observed in MD simulations. The forward model approximates GB migration as a probability distribution over shuffling transformations in the dichromatic pattern. In the forward model, the optimal transformation minimizes net shuffling distance and competes with more costly transformations at finite regularization. Net shuffling cost may be minimized over multiple candidate disconnection modes associated with shear deformations in the dichromatic pattern. Main scientific contributions of our MD survey, in conjunction with analysis from the forward model, are as follows:

\begin{enumerate} 

\item At low and intermediate temperatures, the reorientation geometry of many low $\Sigma$ GBs, including those that shear couple, is dominated by a small set of optimal shuffles which are distance minimizing in the dichromatic pattern. Shear coupling reduces the number of shuffles required for conservative GB migration relative to zero shear transformations. 

\item Changes in microscopic degrees of freedom impact shuffling patterns in significant and quantifiable ways. The boundary plane dependence of shuffling geometry for low $\Sigma$ GBs can be explained by the impact of microscopic translations on distance minimizing mappings in the dichromatic pattern. Small sliding events during migration or changes in GB character may change the dominant shuffling mechanism. 

\item Period doubling effects are discovered in the displacement patterns of GBs with degenerate length shuffles available in the dichromatic pattern, such as the $\Sigma 3$ \hkl(112) GB and the $\Sigma 3$ \hkl(111) GB. In the latter case, degenerate displacements manifest as well known sliding events associated with partial dislocations. 

\item Non-optimal shuffles, often associated with stringlike displacements in the MD data, are found to concentrate in regions where many shuffles of nearly equivalent length can occur (such as the $\Sigma 11$ GB in Figure 16a). When they lie in the plane of the GB, these displacements are associated with GB diffusion.  Non-optimal displacements are common in high $\Sigma$ GBs. 

\item Vortex displacement patterns are ubiquitous during shear coupled motion and tilings of vortices appear to be signatures of constrained migration of \hkl<100> and \hkl<111> symmetric tilt grain boundaries in FCC Ni. The exact combination of mechanisms contributing to these vortex tilings is unclear, with optimal zig-zag motion insufficient to describe all observed displacements. 
 
\end{enumerate}

There are a variety of opportunities to improve upon and extend this work to gain a better scientific understanding of GB migration. One limitation of the displacement texture analysis in this work is the coarse time resolution employed (typically $\Delta t = 0.1$ ns before and after migration). Analysis of characteristic timescales associated with shuffling and dynamical heterogeneity would be beneficial at both shorter and larger migration times to identify rate limiting mechanisms associated with migration and to extend analysis of unit mechanisms to higher temperatures. 

Displacement texture analysis can be extended to other MD datasets to analyze unit mechanisms in complex settings such as curved GB migration, polycrystalline grain growth and GB migration in alloys. Analysis of displacement textures for different metastable GB phases would be particularly interesting to directly probe the impact of microscopic degrees of freedom on shuffling mechanisms. Using the methods in this work, shuffling mechanisms can be systematically enumerated for any CSL GB and crystal structure for varying microscopic shifts and levels of regularization. Both the forward model and displacement texture analysis could help infer migration mechanisms from HRTEM images taken during migration.

\section*{Acknowledgments}
Work at CMU was supported by National Science Foundation grants GFRP-1252522 and DMR-1710186. The authors gratefully acknowledge feedback from  Ankit Gupta, John Hirth, Yuri Mishin, Greg Rohrer, Tony Rollett and Garritt Tucker that helped improve the work. 

\bibliographystyle{elsarticle-num}

\bibliography{citations}

\begin{thebibliography}{10}
\expandafter\ifx\csname url\endcsname\relax
  \def\url#1{\texttt{#1}}\fi
\expandafter\ifx\csname urlprefix\endcsname\relax\def\urlprefix{URL }\fi
\expandafter\ifx\csname href\endcsname\relax
  \def\href#1#2{#2} \def\path#1{#1}\fi

\bibitem{petch1953cleavage}
N.~Petch, The cleavage strength of polycrystals, Journal of the Iron and Steel
  Institute 174 (1953) 25--28.

\bibitem{chiba1994relation}
A.~Chiba, S.~Hanada, S.~Watanabe, T.~Abe, T.~Obana, Relation between ductility
  and grain boundary character distributions in ni3al, Acta metallurgica et
  Materialia 42~(5) (1994) 1733--1738.

\bibitem{randle2010grain}
V.~Randle, Grain boundary engineering: an overview after 25 years, Materials
  science and technology 26~(3) (2010) 253--261.

\bibitem{sangid2013physics}
M.~D. Sangid, The physics of fatigue crack initiation, International journal of
  fatigue 57 (2013) 58--72.

\bibitem{chookajorn2012design}
T.~Chookajorn, H.~A. Murdoch, C.~A. Schuh, {Design of stable nanocrystalline
  alloys}, Science 337~(6097) (2012) 951--954.

\bibitem{shimada2002optimization}
M.~Shimada, H.~Kokawa, Z.~Wang, Y.~Sato, I.~Karibe, Optimization of grain
  boundary character distribution for intergranular corrosion resistant 304
  stainless steel by twin-induced grain boundary engineering, Acta Materialia
  50~(9) (2002) 2331--2341.

\bibitem{rollettrecrystallization}
A.~Rollett, F.~J. Humphreys, G.~S. Rohrer, M.~Hatherly, {Recrystallization and
  related annealing phenomena. 2004}, Elsevier, 2004.

\bibitem{zhang2013stress}
Y.~Zhang, J.~Sharon, G.~Hu, K.~Ramesh, K.~Hemker, Stress-driven grain growth in
  ultrafine grained mg thin film, Scripta Materialia 68~(6) (2013) 424--427.

\bibitem{rupert2009experimental}
T.~J. Rupert, D.~S. Gianola, Y.~Gan, K.~J. Hemker, {Experimental observations
  of stress-driven grain boundary migration}, Science 326~(5960) (2009)
  1686--1690.

\bibitem{li2013incoherent}
N.~Li, J.~Wang, Y.~Q. Wang, Y.~Serruys, M.~Nastasi, A.~Misra, {Incoherent twin
  boundary migration induced by ion irradiation in Cu}, Journal of Applied
  Physics 113~(2) (2013) 23508.

\bibitem{jin2014annealing}
Y.~Jin, B.~Lin, M.~Bernacki, G.~S. Rohrer, A.~Rollett, N.~Bozzolo, Annealing
  twin development during recrystallization and grain growth in pure nickel,
  Materials Science and Engineering: A 597 (2014) 295--303.

\bibitem{lqcke1992texture}
D.~Raabe, K.~L{\"{u}}cke, {Texture and microstructure of hot rolled steel},
  Scripta Metallurgica 26 (1992) 1221--1226.

\bibitem{olmsted2009survey}
D.~L. Olmsted, E.~A. Holm, S.~M. Foiles, {Survey of computed grain boundary
  properties in face-centered cubic metals?II: Grain boundary mobility}, Acta
  Materialia 57~(13) (2009) 3704--3713.

\bibitem{mishin2010atomistic}
Y.~Mishin, M.~Asta, J.~Li, {Atomistic modeling of interfaces and their impact
  on microstructure and properties}, Acta Materialia 58~(4) (2010) 1117--1151.

\bibitem{meiners2020observations}
T.~Meiners, T.~Frolov, R.~E. Rudd, G.~Dehm, C.~H. Liebscher, Observations of
  grain-boundary phase transformations in an elemental metal, Nature 579~(7799)
  (2020) 375--378.

\bibitem{ren2020grain}
X.~Ren, C.~Jin, Grain boundary motion in two-dimensional hexagonal boron
  nitride, ACS nano 14~(10) (2020) 13512--13523.

\bibitem{cahn2004unified}
J.~W. Cahn, J.~E. Taylor, A unified approach to motion of grain boundaries,
  relative tangential translation along grain boundaries, and grain rotation,
  Acta Materialia 52~(16) (2004) 4887--4898.

\bibitem{wei2021direct}
J.~Wei, B.~Feng, R.~Ishikawa, T.~Yokoi, K.~Matsunaga, N.~Shibata, Y.~Ikuhara,
  Direct imaging of atomistic grain boundary migration, Nature Materials (2021)
  1--5.

\bibitem{krause2019review}
A.~R. Krause, P.~R. Cantwell, C.~J. Marvel, C.~Compson, J.~M. Rickman, M.~P.
  Harmer, {Review of grain boundary complexion engineering: Know your
  boundaries}, Journal of the American Ceramic Society 102~(2) (2019) 778--800.

\bibitem{Cantwell2014}
P.~R. Cantwell, M.~Tang, S.~J. Dillon, J.~Luo, G.~S. Rohrer, M.~P. Harmer,
  {Grain boundary complexions}, Acta Materialia 62~(1) (2014) 1--48.
\newblock \href {https://doi.org/10.1016/j.actamat.2013.07.037}
  {\path{doi:10.1016/j.actamat.2013.07.037}}.

\bibitem{zhang2018three}
J.~Zhang, Y.~Zhang, W.~Ludwig, D.~Rowenhorst, P.~W. Voorhees, H.~F. Poulsen,
  Three-dimensional grain growth in pure iron. part i. statistics on the grain
  level, Acta Materialia 156 (2018) 76--85.

\bibitem{zhang2020grain}
J.~Zhang, W.~Ludwig, Y.~Zhang, H.~H.~B. S{\o}rensen, D.~J. Rowenhorst,
  A.~Yamanaka, P.~W. Voorhees, H.~F. Poulsen, Grain boundary mobilities in
  polycrystals, Acta Materialia (2020).

\bibitem{coleman2014effect}
S.~P. Coleman, D.~E. Spearot, S.~M. Foiles, {The effect of synthetic driving
  force on the atomic mechanisms associated with grain boundary motion below
  the interface roughening temperature}, Computational materials science 86
  (2014) 38--42.

\bibitem{priedeman2017role}
J.~L. Priedeman, D.~L. Olmsted, E.~R. Homer, {The role of crystallography and
  the mechanisms associated with migration of incoherent twin grain
  boundaries}, Acta Materialia 131 (2017) 553--563.

\bibitem{bair2019antithermal}
J.~L. Bair, E.~R. Homer, {Antithermal mobility in $\Sigma$7 and $\Sigma$9 grain
  boundaries caused by stick-slip stagnation of ordered atomic motions about
  Coincidence Site Lattice atoms}, Acta Materialia 162 (2019) 10--18.

\bibitem{tucker2011continuum}
G.~J. Tucker, J.~A. Zimmerman, D.~L. McDowell, {Continuum metrics for
  deformation and microrotation from atomistic simulations: Application to
  grain boundaries}, International journal of engineering science 49~(12)
  (2011) 1424--1434.

\bibitem{tucker2012investigating}
G.~J. Tucker, S.~Tiwari, J.~A. Zimmerman, D.~L. McDowell, {Investigating the
  deformation of nanocrystalline copper with microscale kinematic metrics and
  molecular dynamics}, Journal of the Mechanics and Physics of Solids 60~(3)
  (2012) 471--486.

\bibitem{hirth2016disconnections}
J.~P. Hirth, J.~Wang, C.~N. Tom{\'{e}}, {Disconnections and other defects
  associated with twin interfaces}, Progress in Materials Science 83 (2016)
  417--471.

\bibitem{therrien2020minimization}
F.~Therrien, V.~Stevanovi{\'c}, Minimization of atomic displacements as a
  guiding principle of the martensitic phase transformation, Physical Review
  Letters 125~(12) (2020) 125502.

\bibitem{zhang2006characterization}
H.~Zhang, D.~J. Srolovitz, J.~F. Douglas, J.~A. Warren, {Characterization of
  atomic motion governing grain boundary migration}, Physical Review B 74~(11)
  (2006) 115404.

\bibitem{zhang2006simulation}
H.~Zhang, D.~J. Srolovitz, {Simulation and analysis of the migration mechanism
  of {\$}{\$}5 tilt grain boundaries in an fcc metal}, Acta materialia 54~(3)
  (2006) 623--633.

\bibitem{zhang2009grain}
H.~Zhang, D.~J. Srolovitz, J.~F. Douglas, J.~A. Warren, Grain boundaries
  exhibit the dynamics of glass-forming liquids, Proceedings of the National
  Academy of Sciences 106~(19) (2009) 7735--7740.

\bibitem{zhang2007atomic}
H.~Zhang, D.~J. Srolovitz, J.~F. Douglas, J.~A. Warren, {Atomic motion during
  the migration of general [0 0 1] tilt grain boundaries in Ni}, Acta
  materialia 55~(13) (2007) 4527--4533.

\bibitem{mishin2015atomistic}
Y.~Mishin, {An atomistic view of grain boundary diffusion}, in: Defect and
  Diffusion Forum, Vol. 363, Trans Tech Publ, 2015, pp. 1--11.

\bibitem{hirth2019topological}
J.~P. Hirth, J.~Wang, G.~Hirth, {A topological model for defects and interfaces
  in complex crystal structures}, American Mineralogist 104~(7) (2019)
  966--972.

\bibitem{thomas2017reconciling}
S.~L. Thomas, K.~Chen, J.~Han, P.~K. Purohit, D.~J. Srolovitz, {Reconciling
  grain growth and shear-coupled grain boundary migration}, Nature
  communications 8~(1) (2017) 1764.

\bibitem{OT}
I.~Chesser, E.~Holm, B.~Runnels, Optimal transportation of grain boundaries: A
  forward model for predicting migration mechanisms, Acta Materialia 210 (2021)
  116823.

\bibitem{olmsted2009survey2}
D.~L. Olmsted, S.~M. Foiles, E.~A. Holm, {Survey of computed grain boundary
  properties in face-centered cubic metals: I. Grain boundary energy}, Acta
  Materialia 57~(13) (2009) 3694--3703.

\bibitem{ratanaphan2015grain}
S.~Ratanaphan, D.~L. Olmsted, V.~V. Bulatov, E.~A. Holm, A.~D. Rollett, G.~S.
  Rohrer, {Grain boundary energies in body-centered cubic metals}, Acta
  Materialia 88 (2015) 346--354.

\bibitem{foiles2006computation}
S.~M. Foiles, J.~J. Hoyt, {Computation of grain boundary stiffness and mobility
  from boundary fluctuations}, Acta Materialia 54~(12) (2006) 3351--3357.

\bibitem{mendelev2003development}
M.~I. Mendelev, S.~Han, D.~J. Srolovitz, G.~J. Ackland, D.~Y. Sun, M.~Asta,
  {Development of new interatomic potentials appropriate for crystalline and
  liquid iron}, Philosophical magazine 83~(35) (2003) 3977--3994.

\bibitem{plimpton1995fast}
S.~Plimpton, {Fast parallel algorithms for short-range molecular dynamics},
  Journal of computational physics 117~(1) (1995) 1--19.

\bibitem{stukowski2009}
A.~Stukowski, {Visualization and analysis of atomistic simulation data with
  OVITO--the Open Visualization Tool}, Modelling and Simulation in Materials
  Science and Engineering 18~(1) (2009) 15012.

\bibitem{deng2017size}
Y.~Deng, C.~Deng, {Size and rate dependent grain boundary motion mediated by
  disconnection nucleation}, Acta Materialia 131 (2017) 400--409.

\bibitem{yusurvey}
T.~Yu, I.~Chesser, S.~Ratanaphan, E.~Holm, S.~Yang, C.~Deng, Survey of shear
  coupling behavior in fcc ni and bcc fe grain boundaries, Materialia 15
  100945.

\bibitem{chesserout}
I.~Chesser, T.~Yu, C.~Deng, E.~Holm, B.~Runnels, {A continuum thermodynamic
  framework for grain boundary motion}, Journal of Mechanics and Physics of
  Solids 10~(5) (2019) 145--165.

\bibitem{yu2019survey}
T.~Yu, S.~Yang, C.~Deng, {Survey of grain boundary migration and thermal
  behavior in Ni at low homologous temperatures}, Acta Materialia (2019).

\bibitem{Janssens2006}
K.~G. Janssens, D.~Olmsted, E.~A. Holm, S.~M. Foiles, S.~J. Plimpton, P.~M.
  Derlet, Computing the mobility of grain boundaries, Nature materials 5~(2)
  (2006) 124--127.

\bibitem{ulomek2015enfJergy}
F.~Ulomek, C.~J. O'Brien, S.~M. Foiles, V.~Mohles, {Energy conserving
  orientational force for determining grain boundary mobility}, Modelling and
  Simulation in Materials Science and Engineering 23~(2) (2015) 25007.

\bibitem{Trautt2006}
Z.~T. Trautt, M.~Upmanyu, A.~Karma, {Interface Mobility from Interface Random
  Walk}, Science 314~(5799) (2006) 632--635.
\newblock \href {https://doi.org/10.1126/science.1131988}
  {\path{doi:10.1126/science.1131988}}.

\bibitem{schratt2020}
A.~A. Schratt, V.~Mohles, Efficient calculation of the eco driving force for
  atomistic simulations of grain boundary motion, Computational Materials
  Science 182 (2020) 109774.

\bibitem{larsen2016}
P.~M. Larsen, S.~Schmidt, J.~Schi{\o}tz, {Robust structural identification via
  polyhedral template matching}, Modelling and Simulation in Materials Science
  and Engineering 24~(5) (2016) 55007.

\bibitem{MTEX}
F.~Bachmann, R.~Hielscher, H.~Schaeben, Texture analysis with mtex--free and
  open source software toolbox, in: Solid State Phenomena, Vol. 160, Trans Tech
  Publ, 2010, pp. 63--68.

\bibitem{pond2019topological}
R.~C. Pond, J.~P. Hirth, K.~M. Knowles, {Topological model of type II
  deformation twinning in NiTi martensite}, Philosophical Magazine 99~(13)
  (2019) 1619--1632.

\bibitem{hirth1996steps}
J.~P. Hirth, R.~C. Pond, {Steps, dislocations and disconnections as interface
  defects relating to structure and phase transformations}, Acta materialia
  44~(12) (1996) 4749--4763.

\bibitem{pond1983bicrystallography}
R.~C. Pond, D.~S. Vlachavas, {Bicrystallography}, Proceedings of the Royal
  Society of London. A. Mathematical and Physical Sciences 386~(1790) (1983)
  95--143.

\bibitem{runnels2017projection}
B.~Runnels, A projection-based reformulation of the coincident site lattice
  $\sigma$ for arbitrary bicrystals at finite temperature, Acta
  Crystallographica Section A: Foundations and Advances 73~(2) (2017) 87--92.

\bibitem{bilby1965theory}
B.~A. Bilby, A.~Crocker, The theory of the crystallography of deformation
  twinning, Proceedings of the Royal Society of London. Series A. Mathematical
  and Physical Sciences 288~(1413) (1965) 240--255.

\bibitem{bevis1969twinning}
M.~Bevis, A.~Crocker, Twinning modes in lattices, Proceedings of the Royal
  Society of London. A. Mathematical and Physical Sciences 313~(1515) (1969)
  509--529.

\bibitem{christian1995deformation}
J.~W. Christian, S.~Mahajan, Deformation twinning, Progress in materials
  science 39~(1-2) (1995) 1--157.

\bibitem{mompiou2010smig}
F.~Mompiou, M.~Legros, D.~Caillard, {SMIG model: A new geometrical model to
  quantify grain boundary-based plasticity}, Acta Materialia 58~(10) (2010)
  3676--3689.

\bibitem{olmsted2007grain}
D.~L. Olmsted, S.~M. Foiles, E.~A. Holm, {Grain boundary interface roughening
  transition and its effect on grain boundary mobility for non-faceting
  boundaries}, Scripta Materialia 57~(12) (2007) 1161--1164.

\bibitem{holm2010grain}
E.~A. Holm, S.~M. Foiles, {How grain growth stops: A mechanism for grain-growth
  stagnation in pure materials}, Science 328~(5982) (2010) 1138--1141.

\bibitem{chen2020grain}
K.~Chen, D.~J. Srolovitz, J.~Han, Grain-boundary topological phase transitions,
  arXiv preprint arXiv:2011.01475 (2020).

\bibitem{cahn2006coupling}
J.~W. Cahn, Y.~Mishin, A.~Suzuki, {Coupling grain boundary motion to shear
  deformation}, Acta materialia 54~(19) (2006) 4953--4975.

\bibitem{Humberson2019}
J.~Humberson, I.~Chesser, E.~Holm, {Contrasting thermal behaviors in $\Sigma3$
  grain boundary motion in nickel}, Acta Materialia 175 (2019).
\newblock \href {https://doi.org/10.1016/j.actamat.2019.06.003}
  {\path{doi:10.1016/j.actamat.2019.06.003}}.

\bibitem{aditi_science}
A.~Bhattacharya, Y.-F. Shen, C.~M. Hefferan, S.~F. Li, J.~Lind, R.~M. Suter,
  C.~E. Krill~III, G.~S. Rohrer, Grain boundary velocity and curvature are not
  correlated in ni polycrystals, Science 374~(6564) (2021) 189--193.

\bibitem{lu2009strengthening}
K.~Lu, L.~Lu, S.~Suresh, Strengthening materials by engineering coherent
  internal boundaries at the nanoscale, science 324~(5925) (2009) 349--352.

\bibitem{bufford2014situ}
D.~Bufford, Y.~Liu, J.~Wang, H.~Wang, X.~Zhang, In situ nanoindentation study
  on plasticity and work hardening in aluminium with incoherent twin
  boundaries, Nature communications 5~(1) (2014) 1--9.

\bibitem{Thomas2019}
S.~L. Thomas, J.~Han, D.~J. Srolovitz, {The Coupling of Grain Growth and
  Twinning in FCC Metals}, in: IOP Conference Series: Materials Science and
  Engineering, 2019.
\newblock \href {https://doi.org/10.1088/1757-899X/580/1/012026}
  {\path{doi:10.1088/1757-899X/580/1/012026}}.

\bibitem{molodov2011migration}
D.~A. Molodov, T.~Gorkaya, G.~Gottstein, Migration of the $\sigma$7 tilt grain
  boundary in al under an applied external stress, Scripta Materialia 65~(11)
  (2011) 990--993.

\bibitem{wan2017atomistic}
L.~Wan, A.~Ishii, J.-P. Du, W.-Z. Han, Q.~Mei, S.~Ogata, Atomistic modeling
  study of a strain-free stress driven grain boundary migration mechanism,
  Scripta Materialia 134 (2017) 52--56.

\bibitem{bristowe}
R.-J. Jhan, P.~Bristowe, A molecular dynamics study of grain boundary migration
  without the participation of secondary grain boundary dislocations, Scripta
  Metallurgica et Materialia 24~(7) (1990) 1313--1318.

\bibitem{Yan2010}
X.~Yan, H.~Zhang,
  \href{http://www.sciencedirect.com/science/article/pii/S0927025610001515
  http://www.sciencedirect.com/science/article/pii/S0927025610001515/pdfft?md5=c02f15a23b0e6abd5517b085c553f160{\&}pid=1-s2.0-S0927025610001515-main.pdf}{{On
  the atomistic mechanisms of grain boundary migration in \hkl[0 0 1] twist
  boundaries: Molecular dynamics simulations}}, Computational Materials Science
  48~(4) (2010) 773--782.
\newblock \href {https://doi.org/10.1016/j.commatsci.2010.03.029}
  {\path{doi:10.1016/j.commatsci.2010.03.029}}.
\newline\urlprefix\url{http://www.sciencedirect.com/science/article/pii/S0927025610001515
  http://www.sciencedirect.com/science/article/pii/S0927025610001515/pdfft?md5=c02f15a23b0e6abd5517b085c553f160{\&}pid=1-s2.0-S0927025610001515-main.pdf}

\bibitem{mccarthy2020shuffling}
M.~J. McCarthy, T.~J. Rupert, Shuffling mode competition leads to directionally
  anisotropic mobility of faceted $\sigma$11 boundaries in fcc metals, Physical
  Review Materials 4~(11) (2020) 113402.

\bibitem{mccarthy2021alloying}
M.~J. McCarthy, T.~J. Rupert, Alloying induces directionally-dependent mobility
  and alters migration mechanisms of faceted grain boundaries, Scripta
  Materialia 194 (2021) 113643.

\bibitem{schratt2020efficient}
A.~A. Schratt, V.~Mohles, Efficient calculation of the eco driving force for
  atomistic simulations of grain boundary motion, Computational Materials
  Science 182 (2020) 109774.

\end{thebibliography}

\newpage

\appendix
\renewcommand{\thefigure}{S\arabic{figure}}
\setcounter{figure}{0}

\section*{Supplementary Information}

\section{Determination of C-B deformation}\label{sec:appendix_CB_deformation}

In this section we present a compact algorithm for computing the Cauchy-Borne deformation associated with the shear-coupled deformation of a N-atom lattice due to GB motion.
$\{\bm{X}_1,\ldots,\bm{X}_N\}$ and $\{\bm{x}_1,\ldots,\bm{x}_N\}$ are the initial and final positions of atoms in the crystal, respectively, both of which are known in the context of molecular dynamics analysis.
We approximate the corresponding affine deformation gradient as the solution to the following optimization problem
\begin{align}
  \bm{F}^{GB},\bm{u}_0 = \underset{\bm{F},\bm{u}_0}{\operatorname\arg\inf}\frac{1}{p}\sum_i\Big|(\bm{F}\bm{X}_i + \bm{u}_0) - \bm{x}_i\Big|^p
\end{align}
where $p\ge1$.
For the special case of $p=2$, the stationarity condition is
\begin{align}
  \bm{F}^{GB} &= \Big(\sum_m(\bm{x}_m-\bm{u}^0)\otimes\bm{X}_m\Big)\Big(\sum_n\bm{X}_n\otimes\bm{X}_n\Big)^{-1},
  & \bm{u}_0 &= \frac{1}{N}\sum_n(\bm{x}_n - \bm{F}\bm{X}_n).
\end{align}

\begin{proof}
We now switch to index notation.
Subscript indices follow the summation convention and are summed over spatial dimensions.
Superscript indices are only summed explicitly, and correspond to atomic enumeration:
\begin{align}
   \bm{F}^{GB},\bm{u}_0 = \underset{\bm{F},\bm{u}_0}{\operatorname\arg\inf}\frac{1}{2}\sum_n[(F_{pQ}X_Q^n + u^0_p) - x_p^n][(F_{pR}X_R^n + u^0_p) - x_p^n]
\end{align}
The Euler-Lagrange for $\bm{F}^{GB}$ follows by straightforward calculation:
\begin{align}
  0 &\overset{!}{=} \frac{\partial}{\partial F_{iJ}}\frac{1}{2}\sum_n[(\bm{F}_{pQ}X_Q^n + u^0_p) - x_p^n][(\bm{F}_{pR}X_R^n + u^0_p) - x_p^n] \\
  &= \sum_n[(\bm{F}_{pQ}X_Q^n + u^0_p) - x_p^n]\frac{\partial}{\partial F_{iJ}}[(\bm{F}_{pR}X_R^n + u^0_p) - x_p^n]\\
  &= \sum_n[(\bm{F}_{pQ}X_Q^n + u^0_p) - x_p^n]\,[\delta_{ip}\delta_{JR}X_R^n] \\
  &= \sum_n[\bm{F}_{iQ}X_Q^n + u^0_i - x_i^n]\,[X_J^n],
\end{align}
which, when indices are renamed and the expression is  rearranged, produces
\begin{align}
  \bm{F}^{GB}_{iJ}
  =   \Big(\sum_m(x_i^m - u^0_i)\,X_Q^m\Big)\Big(\sum_n\,X_Q^n\,X_J^n\Big)^{-1}
\end{align}
Repeating the process for $\bm{u}$:
\begin{align}
  0 &\overset{!}{=} \frac{\partial}{\partial u^0_i}\frac{1}{2}\sum_n[(\bm{F}_{pQ}X_Q^n + u^0_p) - x_p^n][(\bm{F}_{pR}X_R^n + u^0_p) - x_p^n] \\
    &= \sum_n[(\bm{F}_{pQ}X_Q^n + u^0_p) - x_p^n]\frac{\partial}{\partial u^0_i}[(\bm{F}_{pR}X_R^n + u^0_p) - x_p^n] \\
    &= \sum_n[(\bm{F}_{pQ}X_Q^n + u^0_p) - x_p^n][\delta_{ip}] \\
    &= \sum_n[(\bm{F}_{iQ}X_Q^n + u^0_i) - x_i^n]
\end{align}
Rearranging again produces the result
\begin{align}
  u^0_i = \frac{1}{N}\sum_n(x_i^n - \bm{F}_{iQ}X_Q^n)
\end{align}
The stationarity conditions for $\bm{F}^{GB}$ and $\bm{u}^0$ can be solved iteratively, or can be combined into a single expression depending on the size of the problem.
\end{proof}


\end{document}